\documentclass[letterpaper,11pt]{article}

\usepackage[letterpaper, left=1in, top=1in, right=1in, bottom=1in,nohead,includefoot]{geometry}
\usepackage{color,charter,graphicx,natbib,enumitem,amsmath,amssymb,amsfonts,amsthm,placeins,bm,pdfpages, subfig, placeins,setspace,array,booktabs}
\usepackage[title]{appendix}
\usepackage[svgnames,x11names]{xcolor}
\usepackage[pdftex]
{hyperref}\hypersetup{colorlinks,linkcolor=Black,urlcolor=RoyalBlue3,citecolor=RoyalBlue4}
\usepackage{float}

\usepackage[compact]{titlesec}
    \titlespacing{\section}{0pt}{5pt}{3pt}\titlespacing{\subsection}{0pt}{5pt}{3pt}\titlespacing{\subsubsection}{0pt}{4pt}{3pt}
\titleformat*{\section}{\normalfont\Large\bfseries\blu}
\titleformat*{\subsection}{\normalfont\large\bfseries\blu}
\titleformat*{\subsubsection}{\normalfont\normalsize\bfseries\blu}

\def\eqn#1{eqn.{\,}(\ref{eq:#1})} \def\eqntwo#1#2{eqns.{\,}(\ref{eq:#1},{\,}\ref{eq:#2})} 
\def\parablu#1{\medskip\noindent{\blu\bf #1}}

 \def\seq#1#2{#1{:}#2}
   
\def\diag{\textrm{diag}}
\def\cD{{\cal{D}}}
\def\A{\mathbf{A}}\def\C{\mathbf{C}}\def\F{\mathbf{F}}
\def\G{\mathbf{G}} \def\W{\textrm{W}}
\def\m{\mathbf{m}} \def\R{\mathbf{R}} \def\a{\mathbf{a}}
\def\btheta{{\bm\theta}} \def\bomega{{\bm\omega}} 
\def\bzero{{\bm0}} \def\bI{\mathbf{I}}
\def\tB{\textrm{B}} \def\tE{\textrm{E}} \def\tF{\textrm{F}}
\def\tN{\textrm{N}}  \def\tV{\textrm{V}} \def\tG{\textrm{G}} \def\tT{\textrm{T}}

\def\tiln{{\tilde{n}}}\def\tils{{\tilde{s}}}\def\tilr{{\tilde{r}}}\def\tilq{{\tilde{q}}}

\graphicspath{{./RVDLMfigures/}}

\newcommand{\blind}0 

\if0\blind   
    \def\blu{\color{RoyalBlue4}} 
    \else  
    \def\blu{}  
\fi

\begin{document}\emergencystretch 3em   


\begin{center} 

{\blu\Large Bayesian Dynamic Modeling of Realized Volatility \\ in Financial Asset Price Forecasting} 

\if0\blind
	{ \bigskip
		{\large  
			 Patrick Woitschig\footnote{PhD student, Department of Statistical Science, Duke University, Durham NC 27708-0251 
		   \\ \indent\indent\indent \href{mailto:patrick.woitschig@duke.edu}{patrick.woitschig@duke.edu}}
              \& 
              Mike West\footnote{The Arts \& Sciences Distinguished Professor Emeritus of Statistics \& Decision Sciences   
	     \\ \indent\indent\indent Duke University, Durham, NC 27708, U.S.A. 
		  \href{mailto:mike.west@duke.edu}{mike.west@duke.edu}}}
	} \fi  

\bigskip\bigskip
\today
\bigskip\bigskip

\thispagestyle{empty}\setcounter{page}0

{\blu\bf \large Abstract} 
\end{center} 
We present a new class of Bayesian dynamic models for bivariate price-realized volatility time series in financial forecasting. A novel dynamic gamma process model adopted for realized volatility is integrated with traditional Bayesian dynamic linear models (DLMs) for asset price series. This represents reduced-form volatility leverage and feedback effects through use of realized volatility proxies in conditional DLMs for prices or returns, coupled with the synthesis of higher frequency data to track and anticipate volatility fluctuations.  Analysis is computationally straightforward, extending conjugate-form Bayesian analyses for sequential filtering and model monitoring with simple and direct simulation for forecasting. A main applied setting is equity return forecasting with daily prices and realized volatility from high-frequency, intraday data.  Detailed empirical studies of multiple S\&P sector ETFs highlight the improvements achievable in asset price forecasting relative to standard models and deliver contextual insights on the nature and practical relevance of volatility leverage and feedback effects.  The analytic structure and negligible extra computational cost will enable scaling to higher dimensions for multivariate price series forecasting for decouple/recouple portfolio construction and risk management applications.

\bigskip
\noindent {\em Keywords:} 
Bayesian forecasting, dynamic gamma processes, dynamic linear models,  
leverage effects, risk management, volatility feedback

\newpage 

\section{Introduction}\label{sec:Intro}

We discuss the use of realized volatility and intraperiod data to improve real-time return and risk forecasts. The focus is to extend traditional Bayesian discount dynamic linear models (DLMs) for prices or returns to integrate high-frequency data that  represent lower-noise measures of volatility.  
Our concerns are for efficient sequential filtering and short-term forecasting, with an eye to maximal analytic tractability.   Online systems need methods that learn sequentially, respond quickly as data arrive, and scale across many assets while remaining transparent for routine monitoring.

Existing approaches address parts of this agenda but leave gaps.  Classical stochastic volatility (SV) models of latent log variance processes are often estimated by Markov chain Monte Carlo (MCMC) or particle filtering methods. Such models are flexible but analysis is computationally intensive and requires substantial tuning and monitoring.  Realized volatility (RV) models, such as heterogeneous autoregressive models, treat realized measures as the forecasting target and typically operate outside a joint price--volatility state-space.  Hybrid schemes---realized SV, realized GARCH, and HEAVY---explicitly combine returns with realized measures, 
but they remain either simulation-based or tied to static-likelihood optimization.
Our concern is for analytically tractable, and hence computationally fast and efficient methods to extend existing, well-used models that underpin decouple/recouple portfolio analyses.  

The new class of models introduced here addresses these desiderata. Novel {\em Realized Volatility Dynamic Linear Models} (RV-DLMs) overlay standard Bayesian discount stochastic volatility (SV-DLMs) for prices or returns with a dynamic process for realized volatility based on higher-frequency data. The model component for defined measures of realized variance exploits conditional gamma/inverse gamma theory to deliver fully conjugate Bayesian analyses, with implied scaled-F predictive distributions for realized measures and conditional Student-t distributions  for prices or returns. Sequential filtering, forecasting, marginal-likelihood evaluation, and retrospective smoothing are all performed analytically. The resulting bivariate price--volatility state-space model inherits 
the speed and monitoring ease of discount-learning filters while adding realized measures to improve volatility characterization and short-term prediction.   We extend this framework to incorporate volatility leverage and feedback effects through use of 
realized volatility proxies in conditional DLMs for prices or returns.  The new class of  bivariate price-realized volatility models is able to represent empirically supported effects of  both lagged and contemporaneous high-frequency volatility predictors on price changes, while inheriting the closed-form, fast and efficient Bayesian conjugate analysis of the component DLM and new RV dynamic models.

Section~\ref{sec:background} discusses key stochastic volatility frameworks, both with and without realized volatility inputs, and places our new models in the literature. Section~\ref{sec:joint_prices_vol} reviews traditional Bayesian SV-DLMs, introduces the new class of dynamic volatility models and their integration into RV-DLMs for jointly modeling prices and volatilities, and summarizes key analysis details.   Section~\ref{sec:ETFsApplication} applies these models to a collection of S\&P 500 sector ETFs, comparing RV-DLM variants to the classical SV-DLM benchmark under strict sequential forecasting. Summary comments appear in Section~\ref{sec:conclusion}.

\section{Background and Related Work}\label{sec:background}

 Asset price volatility modeling at short traditional time scales-- such as daily-- has evolved along four broad themes. 
{\em (i) Conjugate discount-learning filtering} based on Bayesian dynamic models that operate sequentially to adapt to 
short-term volatility fluctuations represented by latent volatility states;  these methods are fast, interpretable, and easy to monitor. 
{\em (ii) Simulation-based analyses of more structured stochastic volatility (SV) models} that 
use MCMC or particle filters to capture dynamics at the cost of heavier computation and hands-on tuning.
{\em (iii) Stand-alone realized volatility (RV) models} to forecast empirical measures of high-frequency volatility directly. 
{\em (iv) Joint price-RV systems} that aim to fuse asset returns with realized volatility measures to exploit feedback and measurement information. 
The RV-DLMs introduced here combine the interpretability, analytic speed, and stability of (i) with the empirical sharpness of (iii)–(iv), avoiding the Monte Carlo and tuning burdens typical of (ii).

Traditional Bayesian discount-based approaches down-weight historical information while preserving exact conjugacy that enables efficient sequential analyses. This underlies broadly useful state-space models including those for positive latent scales that underlie the widely used classes of Bayesian dynamic models in financial time series as well as other areas~\citep[e.g.][]{WestHarrison1986,WestHarrison1997,PradoFerreiraWest2021}. 
Classical SV models use a latent log-variance process following an AR(1) evolution~\citep[e.g.][]{Taylor1986,JacquierPolsonRossi1994,Shephard1996}. Inference typically relies on mixture-of-normals approximations within Gibbs sampling~\citep[e.g.][sect.~7.5]{KimShephardChib1998,PradoFerreiraWest2021} or on particle methods~\citep[e.g.][]{West1993a,Aguilar2000,PittShephard1999,Liu2001}. These frameworks are flexible—handling fat tails, leverage, and long memory—but analysis is substantially slower than that based on conjugate discount-learning filters, while monitoring and short-term prediction are typically similar. 

Realized volatility estimates use intraperiod data to define proxies for  integrated variances under idealized microstructure conditions~\citep[e.g.][]{ABDL2003,BNS2002}. Robust estimators address market microstructure noise via subsampling, realized kernels, and related techniques~\citep{BarndorffNielsenHansenLundeShephard2008}. 
Complementary measures based on price range and OHLC (open-high-low-close) data~\citep[e.g.][]{RogersSatchell1991} provide easily computable, low-noise and trend-robust measures popular in practice.  

Hybrid models explicitly combine returns with realized measures. Realized Stochastic Volatility (RSV) augments a standard SV state-space with an RV measurement equation, typically estimated by MCMC or particle filters~\citep[e.g.][]{TakahashiOmoriWatanabe2009,KoopmanScharth2013,TakahashiWatanabeOmori2024,WatanableNakajima2024}. Realized GARCH links a realized measure to the conditional return variance via a measurement equation estimated by (quasi-)maximum likelihood~\citep{HansenHuangShek2012}, while HEAVY models regress daily volatility directly on lagged realized measures~\citep{ShephardSheppard2010}. These hybrids leverage intraday information but still depend on simulation or iterative likelihood optimization.

Our Realized Volatility Dynamic Linear Model (RV–DLM) preserves fully analytic filtering and forecasting by placing a conjugate discount evolution on latent volatilities and modeling realized measures with a multiplicative stochastic process. Marginalizing the latent scale yields closed-form one-step predictive distributions in the scaled-F family. This connects directly to the multiplicative-error model (MEM) tradition: series of positive values are modeled as a latent scale subject to independent, multiplicative shocks over time~\citep{Engle2002,EngleGallo2006}. Then, 
our new model retains the simplicity, analytic tractability and computational ease of Bayesian updating and forecasting in traditional state-space models, with direct interpretability as an extension of such models.  In short, we keep the speed and transparency of conjugate learning but embed realized-measure information so that predictive distributions are calibrated to both high-frequency evidence and direct price fluctuations.

\section{Modeling Framework} 
\label{sec:joint_prices_vol}

The key idea is to extend standard Bayesian dynamic models for equity prices to incorporate additional information arising 
from higher-frequency data, the latter defining {\em realized variance} measurements as additional data. 
We refer to the standard DLM with discount-based stochastic volatility as the SV-DLM, and the new model class as the 
{\em Realized Volatility Dynamic Linear Model} (RV-DLM).     

\subsection{Traditional Bayesian SV-DLM for Price Series} 

The standard discount SV-DLM is the starting point and this subsection summarizes key details and 
notation~\citep[e.g.][sect.~4.3]{PradoFerreiraWest2021}. 
With daily equities as the canonical example, a specific SV-DLM models a 
log price time series $y_t$ over equally-spaced time $t$ via  
\begin{equation}\label{eq:SV-DLM}
\begin{aligned}  y_t &= \F_t' \btheta_t + \nu_t,
  & \nu_t &\sim \tN\bigl(0,v_t\bigr), \quad v_t = 1/\phi_t, 
  \\[4pt]
  \btheta_t &= \G_t \btheta_{t-1} + \bomega_t,
  & \bomega_t &\sim \tN\bigl(\bzero, v_t \W_t\bigr),
\end{aligned}
\end{equation}
with state vector $\btheta_t$ holding time-varying coefficients of elements of the known (at time $t{-}1$) regression vector $\F_t,$ 
time-varying observation variance $v_t$ and implied precision $\phi_t=1/v_t$. The observation and state evolution noise sequences, 
$\{\nu_t\}$ and $\{\bomega_t\}$ respectively, are mutually independent and independent over time, and the state evolution variance matrix
$\W_t$ is known at time $t{-}1.$   The model specification in our
empirical studies is traditional, using $\G_t = \bI$ and \lq\lq small'' $\W_t$ to represent time-varying $\btheta_t$ coefficients. 
The DLM framework is, of course, more general; any linear state-space structure
(time-varying alphas/betas, factor loadings, etc.) can be embedded.

With $\cD_t$ denoting all information-- including the data-- available up to and including time $t,$ the Bayesian analysis has, at time $t{-}1$,  a posterior $p(\btheta_{t-1},\phi_{t-1}|\cD_{t-1})$ of conjugate normal-gamma form: 
 $(\btheta_{t-1}|\phi_{t-1},\cD_{t-1})\sim \tN(\m_{t-1},\C_{t-1}/\phi_{t-1})$ with known $\m_{t-1},\C_{t-1},$ and precision margin  
$(\phi_{t-1}|\cD_{t-1}) \sim \tG(n_{t-1}/2,n_{t-1}s_{t-1}/2)$, a gamma distribution with known shape $n_{t-1}/2$ and rate $n_{t-1}s_{t-1}/2$,
i.e., degrees-of-freedom $n_{t-1}$ and mean $1/s_{t-1}$.
The model is completed with the beta-gamma discount volatility evolution component 
\begin{align}\label{eq:SVmodel}
  \phi_t = \phi_{t-1}\,{\gamma_t}/{\beta}, \quad 
  \gamma_t  \sim\tB\left(n_t^*/2,\; (n_{t-1}-n_t^*)/2\right)  \ \textrm{with}\  n_t^* = \beta n_{t-1},
\end{align}
where the beta distributed innovations $\{\gamma_t\}$  are mutually independent and independent of the $\nu_t,\bomega_t$, and 
the discount factor $\beta\in(0,1]$ controls volatility fluctuations over time
\citep[][sect.~10.8]{WestHarrison1997}.
Following evolution, the time $t$ prior is normal-gamma:  
$(\btheta_{t}|\phi_{t},\cD_{t-1})\sim \tN(\a_{t},\R_{t}/\phi_{t})$  with $\a_t=\G_t\m_{t-1}$ and  $\R_t = \G_t\C_{t-1}\G_t'+\W_t$, and 
$(\phi_{t}|\cD_{t-1}) \sim \tG(n_t^*/2,n_t^*s_{t-1}/2)$
with discounted degrees-of-freedom $n_t^*=\beta n_{t-1}.$   Standard discount factor methods define $\W_t$;  our empirical studies
have $\W_t = \G_t\C_{t-1}\G_t'(1-\delta)/\delta$  for a given state discount factor $\delta\in(0,1]$~\citep[e.g.][sect.~4.3.6]{PradoFerreiraWest2021}.

Forward filtering and forecasting exploits the conjugate structure.  The one-step ahead predictive distribution for $y_t$
is Student-t:  $y_t|\cD_{t-1} \sim \tT_{n_t^*}( f_t,q_t)$ with
$f_t=\F_t'\a_t$ and $q_t = s_{t-1}+\F_t'\R_t\F_t$.  On observing $y_t$ the posterior is normal-gamma: 
 $(\btheta_t|\phi_t,\cD_t)\sim \tN(\m_t,\C_t/\phi_t)$ with $\m_t= \a_t + \A_t e_t$ and $\C_t = \R_t-q_t\A_t\A_t'$ 
 with point forecast error $e_t = y_t-f_t$  and adaptive vector $\A_t = \R_t\F_t/q_t$; the precision margin is 
$(\phi_t|\cD_t) \sim \tG(n_t/2,n_ts_t/2)$ with $n_t=n_t^*+1$ and 
$s_t = r_t s_{t-1}$ where  $r_t = ( n_t^* + e_t^2/q_t )/n_t.$  As $t$ increases, $n_t$ converges to the limiting volatility information level $n_\infty = 1/(1-\beta)$. 

\newpage

\subsection{Dynamic Modeling of Realized Variance}\label{sec:univariateRVmodel}  

Suppose now that there is additional information on the DLM volatility that, specifically, provides 
an unbiased estimate $z_t>0$ of the underlying conditional observation variance $v_t = 1/\phi_t$.   
This may be constructed based on high-frequency returns and overnight prices, for example.  
Traditional such volatility estimates based on averages of squared high-frequency returns naturally suggest model forms involving conditional gamma distributions, while alternative and often more robust proxy realized volatility measures~\citep[e.g.][]{RogersSatchell1991} have conditional distributions that can be approximated by gamma forms. This background suggests the form of the new model introduced here. 

With notation as in the SV-DLM, the {\em realized variance} observation process $\{ z_t \}$ is modeled as  
\begin{equation}\label{eq:ztRVmodel} 
  z_t = {\eta_t}/{\phi_t},\quad  \eta_t \sim\tG({\alpha}/{2},{\alpha}/{2}) 
\end{equation}
where $\alpha>0$. Equivalently,
$ z_t  |  \phi_t \sim\tG({\alpha}/{2},{\alpha}\phi_t/{2}) $ with 
$ \tE[z_t |  \phi_t] = 1/\phi_t = v_t$ 
and $   \tV(z_t |  \phi_t) =2v_t^2/\alpha.$ 
The RV measure $z_t$ is a conditionally unbiased, noisy proxy for the latent variance $v_t$, with
noise level governed by the {\em shape index} hyperparameter~$\alpha$.   

Now suppose that $(\phi_t |\cD_{t-1}) \sim \tG(n_t^*/2,n_t^*s_{t-1}/2)$ with known parameters, as in the standard  SV-DLM. 
Coupling this prior with the structure of the new RV model of~\eqn{ztRVmodel} yields the following results.

\begin{itemize} 
\item  The implied predictive distribution for  $z_t|\cD_{t-1}$ is a scaled-F distribution that can be written as 
$z_t / s_{t-1} \sim \tF(\alpha,n_t^*)$ where $\tF(i,j)$ is a standard F distribution with $(i,j)$ degrees-of-freedom. 
The mean is $ \tE[z_t |  \mathcal D_{t-1}] = s_{t-1} n_t^*/(n_t^*-2)$ if $n_t^*>2.$ 
\item 
The posterior for $\phi_t$ on observing $z_t$ is
$ (\phi_t  |  z_t,\cD_{t-1}) \sim \tG( \tiln_t/2, \tiln_t\tils_t/2)$ 
with $ \tiln_t  = n_t^*+\alpha$ and 
$\tils_t = \tilr_t s_{t-1} $
with  $\tilr_t =  ( n_t^* + \alpha z_t/s_{t-1} )/\tiln_t .$ 
\end{itemize} 
This RV model based on a conditional gamma observation structure for RV estimates $z_t$ is evidently then 
immediately compatible with the conjugate structure of the SV model in the standard SV-DLM.   
Their integration defines the new classes of models, as follows.

\subsection{Bivariate Price--Realized Volatility DLMs: General Structure}
\label{sec:bivarRVDLM}

With notation as in previous sections, the new class of RV-DLMs is defined by combining 
the model structures of~\eqntwo{SV-DLM}{SVmodel} and~\eqn{ztRVmodel} into a bivariate model for $(y_t,z_t).$ 
The conditioning information sets now include past realized variances, so that $\cD_t = \{ y_t, z_t, \cD_{t-1} \}$ for all $t\ge 0.$
The new observation level model has general compositional  form
\begin{equation}\label{eq:bivarRVDLM} 
 p(y_t,z_t | \btheta_t,\phi_t, \mathcal D_{t-1})
  =
  p(y_t |z_t,\btheta_t,\phi_t,\cD_{t-1}) p(z_t  |  \phi_t,\cD_{t-1}) 
\end{equation} 
with the two component pdfs those of the models of~\eqntwo{SV-DLM}{ztRVmodel}. 
The generality of the model class comes through the opportunity to introduce dependence on $z_t$ of the first component pdf here, that from the DLM for $y_t.$ Our specific examples  do this by including functions of $z_t$ as linear predictors: the DLM regression vector $\F_t$ includes one or more functions of $z_t$ as well as their lagged values (that form part of $\cD_{t-1})$ and their time-varying coefficients are elements of $\btheta_t.$

For the states $(\btheta_t,\phi_t),$ the evolution model components 
and implied conjugate normal-gamma prior:posterior distributions are as in the standard SV-DLM
but now with modified defining hyperparameters.  With notation as in the SV-DLM,  
the new model of~\eqn{bivarRVDLM} coupled with the prior defined by
 $(\btheta_t|\phi_t,\cD_{t-1})\sim \tN(\a_t,\R_t/\phi_t)$ and
$\phi_{t}|\cD_{t-1} \sim \tG(n_t^*/2,n_t^*s_{t-1}/2)$ yields the following. 

\parablu{One-Step Forecasting:} 
The one-step predictive distribution for $(y_t,z_t|\cD_{t-1})$  exhibits dependencies through that of $\F_t$ on $z_t$ as well as via scale dependencies induced through the common latent volatility process $\{\phi_t\}.$   
The joint forecast pdf can be expressed in the compositional form
\begin{equation}\label{eq:bivarRVDLMonestep} 
 p(y_t,z_t | \mathcal D_{t-1}) 
  =
  p(y_t |z_t,\cD_{t-1}) p(z_t  | \cD_{t-1}) 
\end{equation} 
where: (i) the margin for $z_t$ is scaled-F as in Section~\ref{sec:univariateRVmodel}; and (ii) the conditional for $y_t$ at any $z_t$ is Student-t as in the SV-DLM, now with location and scale parameters depending on $z_t.$ The latter is  
$y_t|z_t,\cD_{t-1} \sim \tT_{\tiln_t}( f_t,\tilq_t)$ with
$f_t=\F_t'\a_t$ and $\tilq_t = \tils_t+\F_t'\R_t\F_t$, both dependent on $z_t$ in cases where functions of $z_t$ are chosen as 
predictors in $\F_t.$ 

Technically, forecasting is computationally trivial via direct compositional simulation of the predictive distribution of $(y_t,z_t|\cD_{t-1})$. A single draw from the bivariate predictive is generated by simulating $z_t$ from the 
scaled-F distribution $p(z_t|\cD_{t-1})$ and then, conditional on that synthetic value, simulating $y_t$ from the Student-t $p(y_t|z_t,\cD_{t-1}).$

\parablu{Filtering Updates:}
Conditional on observing $z_t$ alone,  the precision margin is updated as in Section~\ref{sec:univariateRVmodel}; that is,  
$ (\phi_t  |  z_t,\cD_{t-1}) \sim \tG( \tiln_t/2, \tiln_t\tils_t/2)$ 
with $ \tiln_t  = n_t^*+\alpha$ and 
$\tils_t = \tilr_t s_{t-1} $
where  $\tilr_t =  ( n_t^* + \alpha z_t/s_{t-1} )/\tiln_t .$ 
  
On then also observing $y_t,$  we have $\cD_t = \{ y_t, z_t, \cD_{t-1}\}.$  The filtering update yields the normal-gamma posterior with state conditional
 $(\btheta_t|\phi_t,\cD_t)\sim \tN(\m_t,\C_t/\phi_t)$  precisely as in the SV-DLM but now noting the dependence on $z_t$ through $\F_t$. Also, the prior 
 for $\phi_t$ has already been updated to condition on $z_t$ so that the resulting 
precision margin is correspondingly modified relative to that in the SV-DLM. Specifically, 
$\phi_t|\cD_t \sim \tG(n_t/2,n_ts_t/2)$ with 
$n_t=\tiln_t+1 = n_t^* + \alpha+1$  and 
$s_t = r_t \tils_t = r_t\tilr_t s_{t-1}$ where now $r_t = (\tiln_t + e_t^2/\tilq_t)/n_t$.

The RV model shape index $\alpha$ plays the role of an effective degrees-of-freedom parameter in learning about the latent $\phi_t$ process:  the information content for $\phi_t$ of the single price observation $y_t$ is 1 relative to $\alpha$  from the realized variance $z_t.$ 
As $t$ increases, the degrees-of-freedom
$n_t$ converges to the limiting information level $n_\infty = (1+\alpha)/(1-\beta)$.

\bigskip 
 This bivariate construction neatly couples realized volatility information to prices while retaining
analytic, conjugate filtering and forecasting structure.  
This allows  flexibility in modeling the relationship between prices and realized volatilities through choices of how $z_t$-- and the past outcomes $z_{t-j}\in \cD_{t-1}$ for $j\ge 1$--  enter into the DLM defining $p(y_t |z_t,\btheta_t,\phi_t,\cD_{t-1}).$   One subclass of models 
is based on assuming conditional independence of  $y_t$ and $z_t$ given the latent state vector $\btheta_t$ in the price DLM; 
then the latent volatility process $\phi_t$  defines the only  linkages between the two data series.  This assumption is certainly relevant as an approximation in some applications, but may be questioned in others. The following section addresses a more interesting and practically relevant class of RV-DLMs.

\subsection{Volatility Leverage and Feedback Effects\label{sec:leverage}} 

Empirically observed relationships between prices and volatility are often subtle and more complicated than represented purely 
through the latent precision process $\phi_t$.  In particular, studies of stochastic 
volatility (SV) models~\citep[e.g.][]{OMORI2007425,NAKAJIMA20092335,NAKAJIMA20123690} have addressed leverage and feedback effects.
High-frequency evidence is especially relevant here. \citet{BollerslevLitvinovaTauchen2006} use five-minute S\&P 500 futures returns to distinguish lead--lag patterns, documenting negative associations between absolute high-frequency returns and current/past returns, negligible reverse dependence from lagged volatility to future returns, and a strong contemporaneous relation that can be interpreted as near-instantaneous volatility feedback. Their results reinforce that daily price--volatility dependence is best treated as a reduced-form combination of lagged and contemporaneous channels rather than as a clean structural split.
Such effects are particularly relevant for individual stocks and sector funds, where volatility shocks can be sector-specific. The new RV-DLM framework allows for aspects of such dependence by exploiting the flexibility of the DLM component to integrate information from the RV estimates into the predictive regression for prices.  
 
First, the regression vector $\F_t$ in the SV-DLM component of \eqn{SV-DLM} can include elements that are functions of the contemporaneous RV $z_t$; this is enabled by the compositional form of the bivariate model specification in \eqn{bivarRVDLM}.  Second, other 
elements of $\F_t$ can  represent predictive information from past RV outcomes, $z_{t-j}$ for $j>0$ as they are part of the historical information set $\cD_{t-1}.$   In our empirical case study of log prices for 
multiple S\&P ETFs,  we have models that include 
$x_t = \sqrt{z_t}$, the realized standard deviation (SD), as a contemporaneous predictor, and $x_{t-1}$ as a lagged predictor. 
The models also include a time-varying intercept and time-varying AR(1) term, i.e., lag$-1$ log price $y_{t-1}$ in $\F_t,$ so that the realized volatility predictors represent excess predictive value for changes in asset prices over time, with easy access to inferences on time-varying coefficients and the resulting regression effects. 

With current and lag$-$1 realized volatility predictors of price changes, the model represents price--volatility dependence through the pair of dynamic coefficients on $x_t$ and $x_{t-1}$. Conditional on past volatility, the contemporaneous coefficient measures the immediate association of realized SD with the current log-price nowcast, while the lagged coefficient measures the short-horizon effect of recent realized SD on the next price forecast. The relevant empirical signal is the net realized-volatility contribution, say $\theta_{0x,t}x_t+\theta_{1x,t}x_{t-1}$, so leverage and feedback interpretations depend on the relative signs, magnitudes and posterior dependence of these coefficients. In this sense, the RVL-DLM is best viewed as a reduced-form daily representation of contemporaneous and lagged return--volatility asymmetries, not as a structural separation of the two causal mechanisms. This provides a flexible representation of classical leverage intuition~\citep{black_1976} and asymmetric volatility responses~\citep[e.g.][]{engle_ng_1993}; the sign of each effect is learned from the data rather than imposed through explicit GARCH-type asymmetry terms. Also, as $\phi_t$ simultaneously governs the innovation variance and is updated by both $z_t$ and the magnitude of price forecast errors, large adverse price moves propagate into subsequent variance learning through the sequential filter.

In principle, the model could be extended to include more or different 
functions of the realized-volatility proxy $z_t$ and its lags to more elaborately model non-linear effects, though we restrict attention to this first specification in our studies presented here.      

\subsection{Retrospective Smoothing}

The conjugate filtering analysis also supports retrospective analysis over any fixed interval $1{:}T$.  At each time $t$, the filtered posterior has the same normal-gamma form as in the standard SV-DLM, with the RV observation absorbed through the updated gamma parameters $(n_t,s_t)$.  Fixed-interval smoothing is therefore based on the usual backward recursion 
$$
  p(\btheta_t,\phi_t|\cD_T) = \int p(\btheta_t,\phi_t| \btheta_{t+1},\phi_{t+1},\cD_t)
  p(\btheta_{t+1},\phi_{t+1}|\cD_T)\,d\btheta_{t+1}d\phi_{t+1},
$$
with the same backward updating and backward sampling logic as in traditional discount SV-DLMs. Thus smoothed state estimates, smoothed volatility estimates and simulated retrospective trajectories require only minor modifications of standard DLM algorithms~\citep[][chap.~4]{PradoFerreiraWest2021}.

\section{Case Study: S\&P Sector ETFs}    \label{sec:ETFsApplication}
\subsection{ETF Data and Volatility} \label{sec:ETFdatamodels}

Empirical studies focus on daily prices for the 9 Select Sector SPDR ETFs with full availability over the study period, together with the S\&P Index itself. The sectors are taken from the set of S\&P Select SPDR Funds, named and labeled in Table~\ref{tab:ETFtableInfo}.  
\begin{table}[ht!]
\centering
\begin{tabular}{llll}
XLB: Materials  & XLE: Energy           & XLF: Financial     & XLI: Industrial  \\
XLK: Technology & XLP: Staples & XLV:  Health Care  & XLY: Discretionary  \\
& XLU: Utilities   & S\&P:  S\&P Index & \\
\end{tabular}
\caption{S\&P sector ETFs and Index.}
\label{tab:ETFtableInfo}
\end{table}

Raw data are sourced from \href{https://finance.yahoo.com/}{Yahoo Finance}. This 
provides daily ETF price information including open, high, low, and 
closing prices designated $O_t,H_t,L_t,C_t$  respectively on each day $t$. 
After using the first trading day to define lagged predictors, the 
modeled observations span 1/4/2000--12/31/2025, a sample of 6,538 trading days 
covering a range of financial-economic regimes including the GFC and Covid-19 
pandemic.

We take $y_t$ as log price rather than \lq\lq differencing away'' potentially useful information by modeling returns. This is our preferred approach that allows for opportunities to improve short-term predictions via structured models for prices directly~\citep[as used, for instance,  recent examples in][]{TallmanWest2024}. 
That is, the price series is $y_t =\log(C_t)$.   For the RV measure $z_t$ we adopt  the  Rogers--Satchell realized variance~\citep{RogersSatchell1991} defined by $z_t=  \log(H_t/C_t)\log(H_t/O_t) + \log(L_t/C_t)\log(L_t/O_t).$ 
This OHLC--based metric is drift-robust relative to alternatives, and can be directly evaluated on the limited high-frequency price information available. Theoretical background and relationships with integrated high-frequency volatilities underpin the selection of this measure and the relevance of our conditional RV gamma process model~\citep{RogersSatchell1991}.

\begin{figure}[t]
  \centering
    \includegraphics[width=0.495\linewidth]{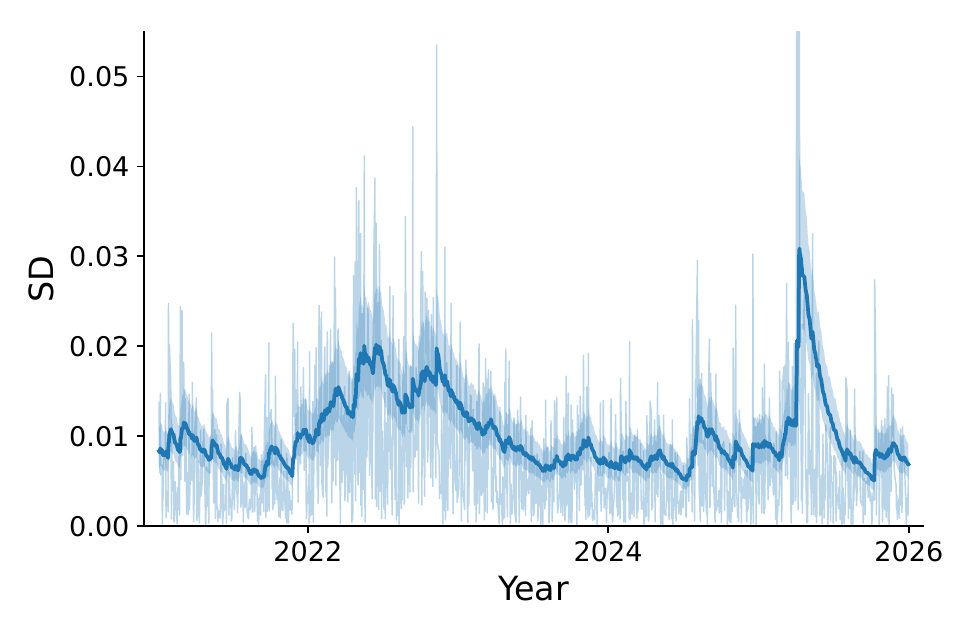}
    \includegraphics[width=0.495\linewidth]{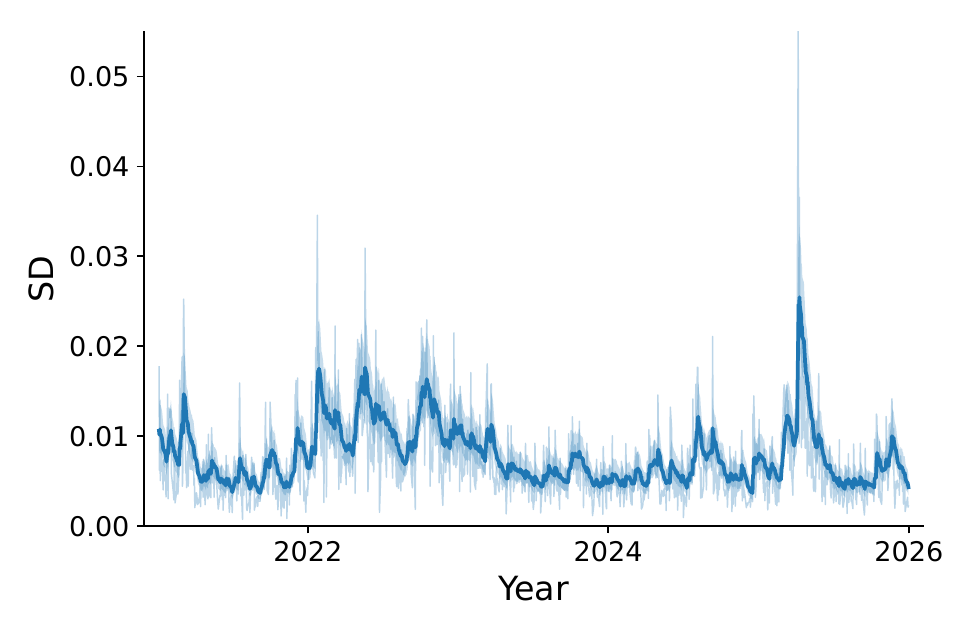}
   \caption{Trajectories of inferences on the latent observation SD
  $\sqrt{v_t}=1/\sqrt{\phi_t}$ in the S\&P Index under two  models. On each day, these show medians (solid line) and equal-tailed 90\% credible intervals (shaded)  of filtered posterior based on data up to that day. 
  {\em Left:} Based on the price series discount volatility analysis using daily log-price changes alone.
  {\em Right:} Based on the dynamic gamma RV model for the realized variance proxy $z_t$ alone.
  }
  \label{fig:BOSSslidefigure}
\end{figure}

The relevance of realized volatility information is highlighted in
Figure~\ref{fig:BOSSslidefigure}. The figure compares filtered trajectories
of posterior inferences on the latent observation standard deviation
$\sqrt{v_t}=1/\sqrt{\phi_t}$ from two separate analyses of the S\&P Index
series. The figures show trajectories over just a few recent years simply to highlight the concordance and complementarity of information on volatility fluctuations.  The first is from a price series discount-volatility analysis based solely on daily log-price changes, with no realized-volatility input.  The second is from the simple dynamic gamma model
for the realized variance proxy $z_t$ alone. These analyses use the same
hyperparameter specifications as those adopted in the ETF
studies of Section~\ref{sec:ETFmodelstrainpriorsparams}. In each case, the
sequential analysis yields posterior gamma distributions
$p(\phi_t|\cD_t)$ over time; these define the posterior
medians and  credible intervals for $\sqrt{v_t}$. The main
points are that: (i) the two separate analyses---one based only on the price
series and the other only on the realized-volatility series---produce
similar trajectories of inferred volatility over time; and (ii) the RV analysis yields more precise trajectories, indicating additional precision gained from the intraday data.  This  empirically supports the relevance of realized-volatility data as
information on the latent observation variance, underpinning its
integration with the price DLM in the joint RV-DLM framework.

 \newpage
 
\subsection{Models, Priors and Hyperparameters} \label{sec:ETFmodelstrainpriorsparams}

We separately analyze each ETF series and the S\&P Index  with each of  3 models, making comparisons across results to elucidate features of the analysis.
The most general model for each ETF has the DLM component for log prices with a local intercept, an AR(1) term, and volatility predictors 
$x_t$ and $x_{t-1}$ where $x_t=\sqrt{z_t}$ is realized SD on day $t$.   The corresponding coefficients in $\btheta_t$ are all potentially time-varying. 
We take $\G_t=\bI$ so that time-variation in these coefficients is allowed as a random walk, with variation controlled by the DLM discount factor $\delta.$ 
Refer to this model as the RVL-DLM, the \lq\lq L'' in RVL now representing the conventional leverage/instantaneous-feedback terminology associated with contemporaneous price--volatility dependence through $x_t$.  Lagged volatility information is represented through the inclusion of lagged RV $x_{t-1}$ in each of the 3 models. For each ETF we consider and compare: 
\begin{itemize} \itemsep-2pt
\item the bivariate  RVL-DLM for $(y_t,z_t)$ noted above, with $\F_t = (1, y_{t-1}, x_t, x_{t-1} )'$;
\item the simpler bivariate RV-DLM with $\F_t = (1, y_{t-1}, x_{t-1} )'$, relying on the lag$-1$ RV predictor $x_{t-1}$ but not involving the contemporaneous value $x_t$; 
\item the univariate SV-DLM for $y_t$ alone that has the same reduced $\F_t = (1, y_{t-1}, x_{t-1} )'$. 
\end{itemize}
The benchmark in these comparisons is a traditional SV-DLM of the same operational Bayesian forecasting class. In the third model, lagged realized volatility enters only as a known predictor in the price DLM. Unlike the RV-DLMs, this model does not include a realized-volatility observation equation and hence does not use $z_t$ for direct sequential learning about $\phi_t$. The methodological focus is on extending this operational Bayesian forecasting class in a fully  conjugate sequential form, rather than on conducting an exhaustive comparison with other models. 
Comparing these three  models provides a focused assessment of the incremental value of realized-volatility information in the SV-DLM framework: through the lagged realized volatility predictor, and then  through the additional contemporaneous realized volatility predictor that defines the RVL-DLM component. The comparison therefore addresses the central question of whether  realized volatility, incorporated in a fully conjugate sequential model, improves short term price forecasting while retaining closed-form filtering, forecasting, and marginal-likelihood  evaluation.

We split the data into an initial pre-2010 period of {2,514} trading days over 1/4/2000--12/31/2009, and a follow-on evaluation period of {4,024} trading days over {1/4/2010--12/31/2025}. The first trading day in the sample is used to define lagged predictors, so the first modeled observation enters on {1/4/2000}. The post-2010 evaluation period begins on the first trading day after the scoring start date at {1/4/2010}. 
For each ETF and each of the 3 models, analyses are based on pre-specified hyperparameters chosen based on prior analysis of preceding data.  In real-time application, hyperparameters can be chosen based on marginal likelihood evaluations or more customized methods and adapted over time; for the examples here, that is not a main focus and pre-specified values serve our purposes.  Specifically, the SV-DLM uses $(\delta,\beta)= {(0.999,0.925)}$, while both the RV-DLM and RVL-DLM use $(\delta,\beta,\alpha)={(0.999, 0.875, 2.75)}$. These values of $\delta,\beta$ are consistent with their specifications for daily sector and index change data alone, prior to consideration of realized volatility information.  The chosen $\alpha=2.75$ is based on marginal likelihood evaluations over the training data period.  For each sector series and the index itself, these suggest and support values in the range {2.5--3.5}. Customization to select $\alpha$ as series-specific is of course relevant in real-time application. For the neutral comparison with the RV-DLM here, we chose a consensus value to apply in analysis of each of the series.  Note that the value $\alpha=2.75$ indicates that the RV data provide almost three times the information on the volatility process $\phi_t$ as that provided by the return series alone in the context of uncertainties about the DLM model states.  Then,  inferences on the state vector $\btheta_t$ are only modestly impacted by  the value of $\alpha$ in terms of location of inferred values, though of course the feed-through from the RV data influences the precision with which the values are inferred. The value of $\alpha$ does, of course, materially impact on the precision of forecast distributions and this plays a key role in assessing predictive performance.
Following the training data analysis, the values of these 3 hyperparameters are held fixed for the following sequential filtering and forecasting analysis over {2010--2025}. That is, there is no further tuning, re-optimization, or rolling recalibration for the purposes of strictly out-of-sample model evaluation and comparison in this particular study.

Initial priors at the start of the pre-2010 training period are relatively vague with $\a_1={(0,1,0)'}$ and $\R_1= {\diag(0.10,0.01,0.05)/0.999}$ for the SV-DLM and RV-DLM, and with $\a_1= {(0,1,0,0)'}$ and $\R_1= {\diag(0.10,0.01,0.05,0.05)/0.999}$ for the RVL-DLM. 
Initial variance information is $n_1^*=\beta$ to define a vague initial prior
in all models. 
The ETF-specific initial scale $s_1$ is set to a prior point estimate chosen to be consistent with pre-2000 data for the corresponding series.  These pre-sample values are used only to define the initial prior scale and are not included in the modeled sample or in any predictive-score evaluation.  Forward filtering through the pre-2010 sample then serves as the warm-up analysis that defines the posteriors carried into {1/4/2010}.

Following the pre-2010 filtering period, the posteriors for model states at {12/31/2009} in the corresponding DLMs evolve one trading day to define the priors for  {1/4/2010}. At that point, marginal likelihood computations are re-initialized to deliver cumulative out-of-sample predictive scores for each of the 3 models for each ETF over the following post-2010 evaluation period. These cumulative log predictive density scores are the key measures of comparative out-of-sample predictive performance over this period.

\begin{figure}[t!]
  \centering
  \centering
    \includegraphics[width=0.495\textwidth]{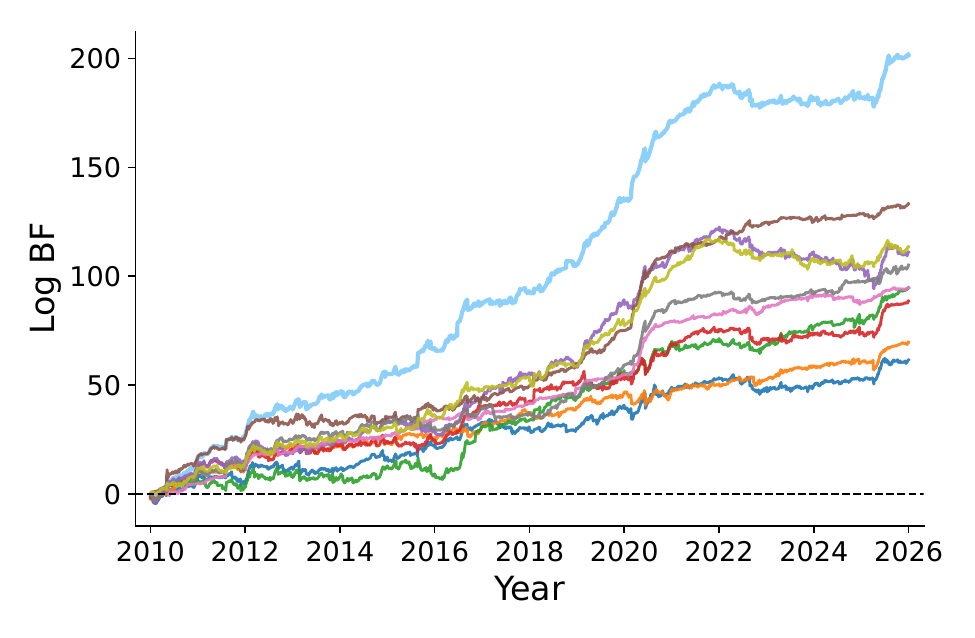}
    \includegraphics[width=0.495\textwidth]{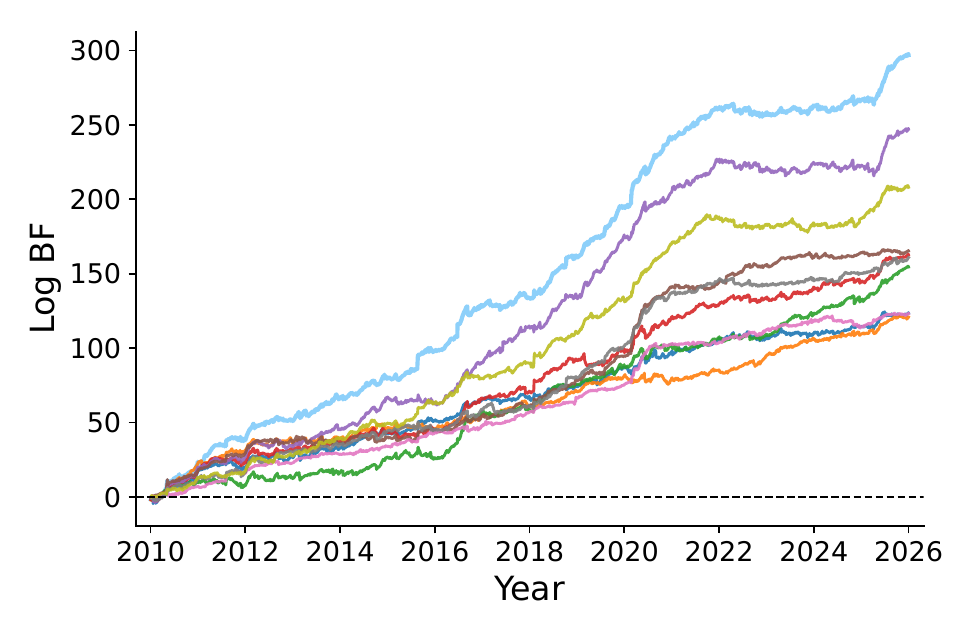}    \includegraphics[width=0.495\textwidth]{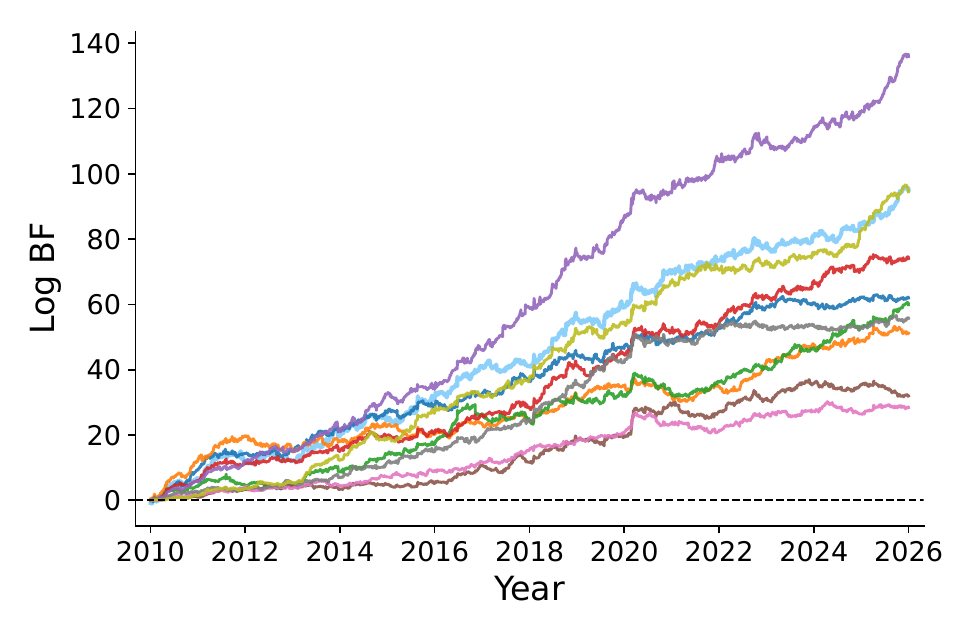}
  \caption{Cumulative log Bayes factors over the post-2010 period for each of the 9 ETFs and the S\&P Index series. 
  {\em Upper left:} Evidence in favor of RV-DLM over SV-DLM;
  {\em upper right:}  Evidence in favor of RVL-DLM over SV-DLM;
  {\em lower:}  Evidence in favor of RVL-DLM over RV-DLM.
  The zero line is indicated in each frame.  Two cases  for which the RVL-DLM is most dominant are 
  the S\&P Index itself (light blue trajectory) and XLK, the Technology sector ETF (purple trajectory).}
  \label{fig:cumulativeLogBFall}
\end{figure}

\begin{figure}[t!]
  \centering
    \includegraphics[width=0.495\textwidth]{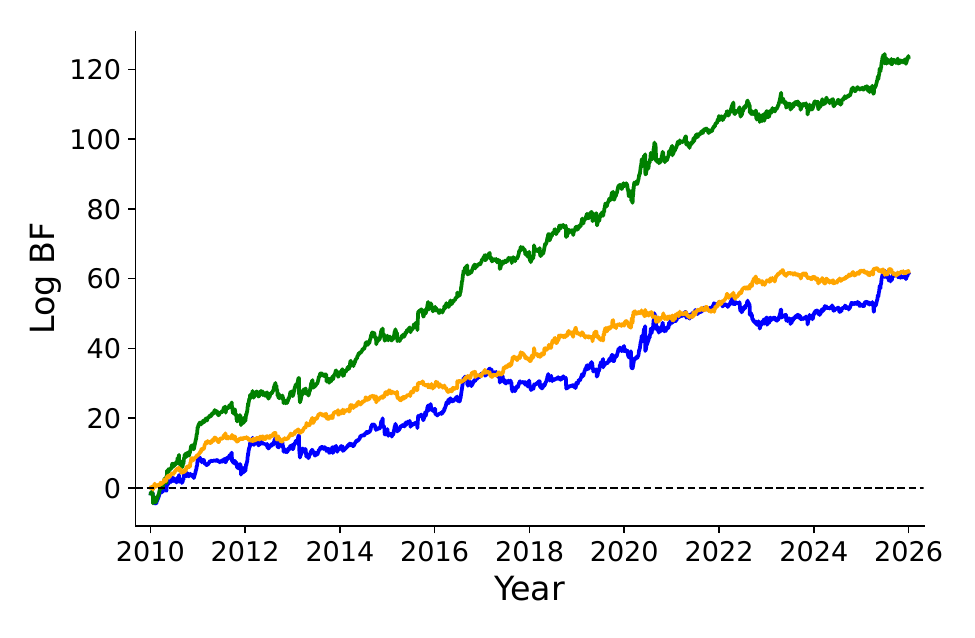}
    \includegraphics[width=0.495\textwidth]{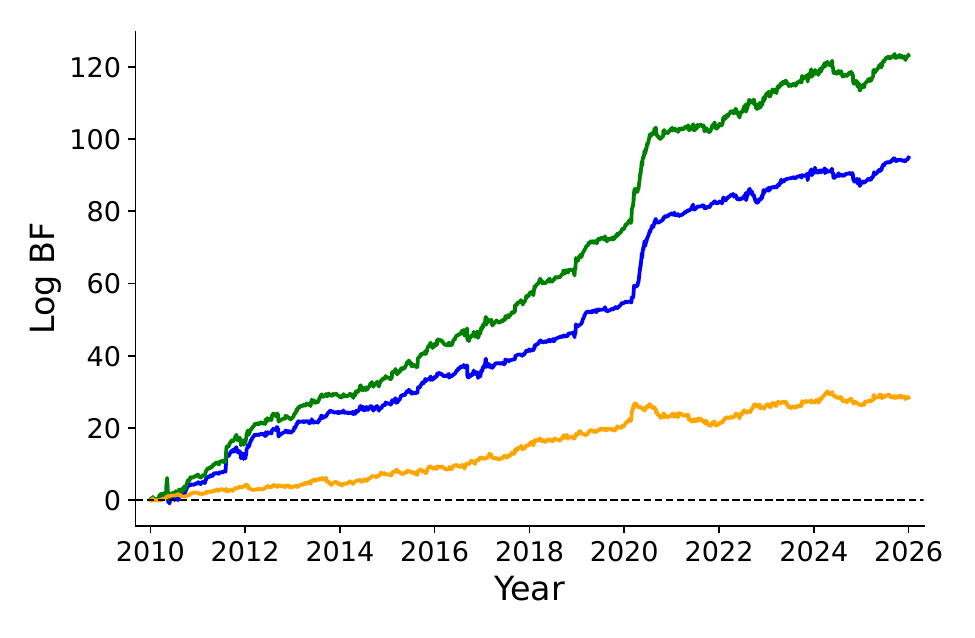}
    \includegraphics[width=0.495\textwidth]{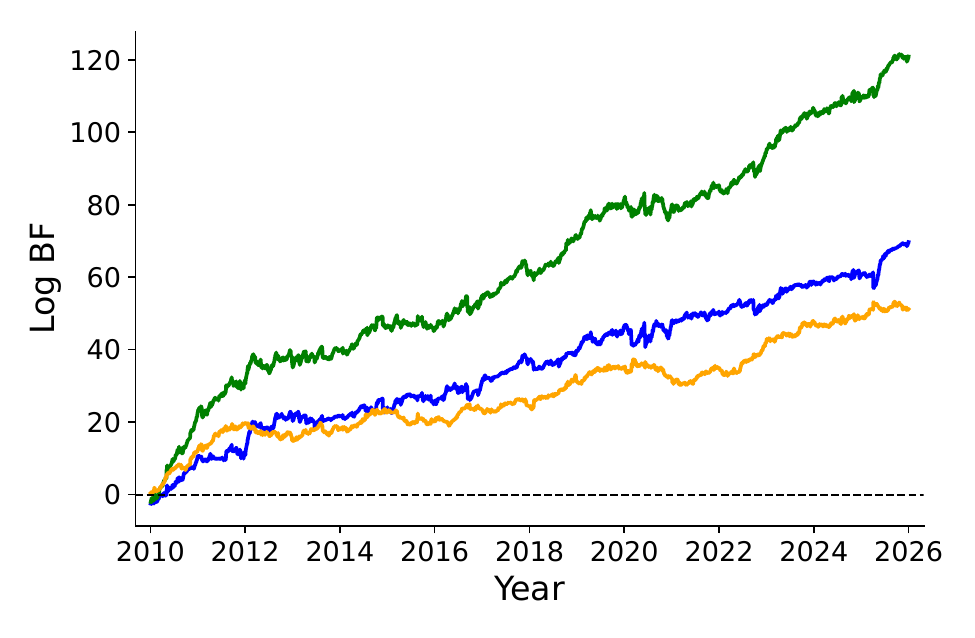}
  \caption{Cumulative log Bayes factors over the post-2010 period for 3 of the ETFs: {\em upper left:} XLB Materials; {\em upper right:} XLU Utilities; {\em lower:} XLE Industrial.
  The zero line is indicated.
  Blue trajectories represent evidence in favor of RV-DLM over SV-DLM, the green that for RVL-DLM over SV-DLM, and the orange that in favor of RVL-DLM over RV-DLM.}
  \label{fig:cumulativeLogBF3ETFs}
\end{figure}
 
\newpage

\subsection{Post-2010 Evaluation and Inferential Summaries}\label{sec:post2010eval}

\subsubsection{Predictive Model Scores}\label{sec:logBFtrajectoriesforETFs}

Figures~\ref{fig:cumulativeLogBFall} and~\ref{fig:cumulativeLogBF3ETFs} show trajectories over time of the log Bayes factors comparing pairs of models for prediction of prices. Index models by 
$m\in\{\textrm{SV-DLM, RV-DLM, RVL-DLM}\}$. For the log price series $y_t$ on any one ETF, the Bayes factor in favor of model $m$ relative to another model $m'$ is simply $B_t(m,m') = p(y_{\seq 1t}|\cD_0,m)/p(y_{\seq 1t}|\cD_0,m').$    As data accumulates, this updates 
via $B_t(m,m') = B_{t-1}(m,m')  p(y_t|\cD_{t-1},m)/p(y_t|\cD_{t-1},m'),$ i.e.,  $\log( B_t(m,m'))$ cumulates over time by adding relative one-step predictive log pdf \lq\lq scores''.      Figure~\ref{fig:cumulativeLogBFall} shows trajectories of these log scores for $(m,m')$ pairs (RV-DLM, SV-DLM), 
(RVL-DLM, SV-DLM) and (RVL-DLM, RV-DLM) and all ETF series.  This shows persistently growing evidence in favor of the 2 RV-DLM model variants over the traditional SV-DLM for all ETFs and the S\&P Index itself. 
Figure~\ref{fig:cumulativeLogBF3ETFs} highlights this improved predictive performance of the RV-DLM models for 3 of the ETFs. For these selected sectors--XLB, XLU and XLE--the same ordering seen in Figure~\ref{fig:cumulativeLogBFall} is evident: both RV model variants dominate the SV-DLM, and the RVL-DLM further improves on the RV-DLM. Thus, in the current analysis, the contemporaneous realized-volatility predictor adds positive cumulative predictive value beyond the lagged-RV specification across the ETF and Index series, though the size and timing of the gain varies by sector.

A detail notable in several of the ETF analyses is exemplified by XLU-- representing the Utilities industry sector; there is a substantial jump in Bayes factors in favor of the RV model variants in the early weeks of 2020, indicative of the specific impact of the realized volatility information in forecasting price fluctuations at the start of the Covid-19 pandemic. Otherwise, a main conclusion is that the RV-DLM model variants substantially improve short-term predictions for all ETFs on the basis of this traditional predictive score. Then, the model with the contemporaneous realized-volatility component is superior across all ETF and Index series in this evaluation. The incremental RVL-DLM gain over RV-DLM is smaller than the gain of either bivariate model over SV-DLM, but it is persistent and cumulatively material. Practical differences between the two RV model variants in their inferences on state vectors and volatilities over time-- and hence on short-term predictions-- are modest. There are minor differences in point forecasts with slight improvements under RVL-DLM, while this model gains more in terms of relatively reduced predictive uncertainties and sharper one-step predictive densities.

\subsubsection{Realized Volatility Predictor Coefficients: Leverage and Feedback}\label{sec:ThetatrajectoriesforETFs}

\begin{figure}[p!]
  \centering
    \includegraphics[width=0.495\textwidth]{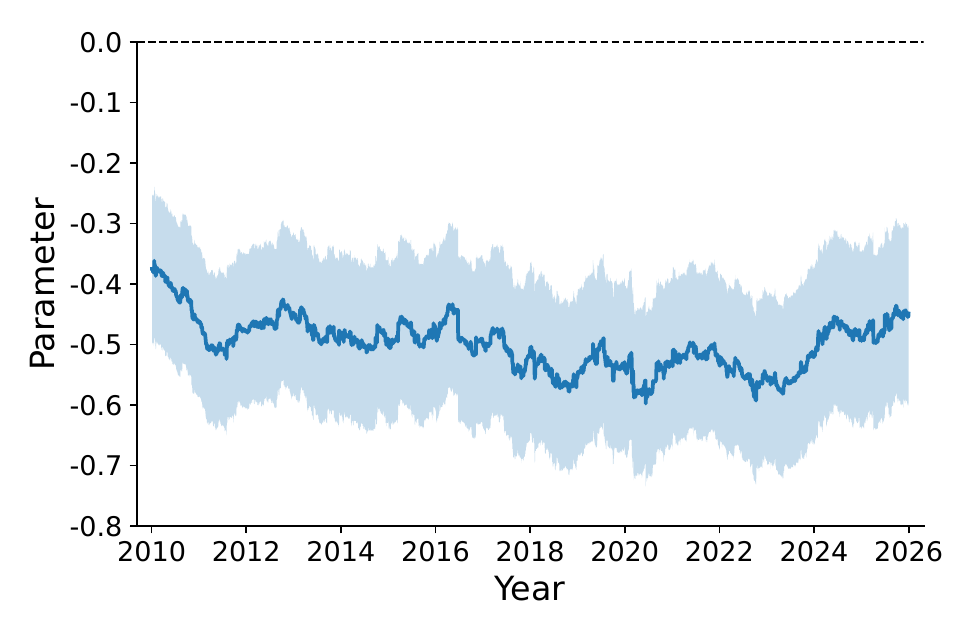}
    \includegraphics[width=0.495\textwidth]{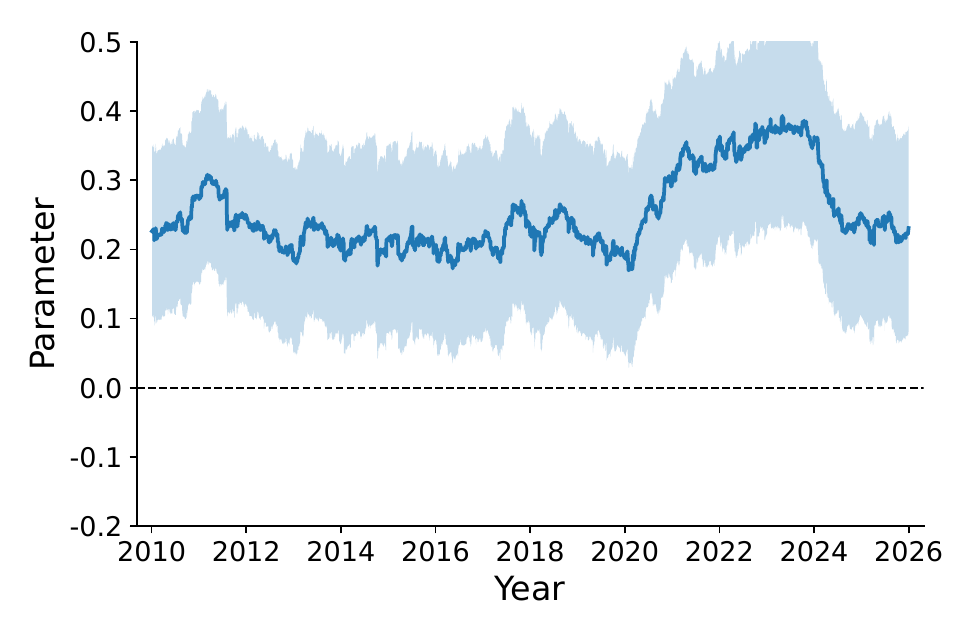}
    \includegraphics[width=0.495\textwidth]{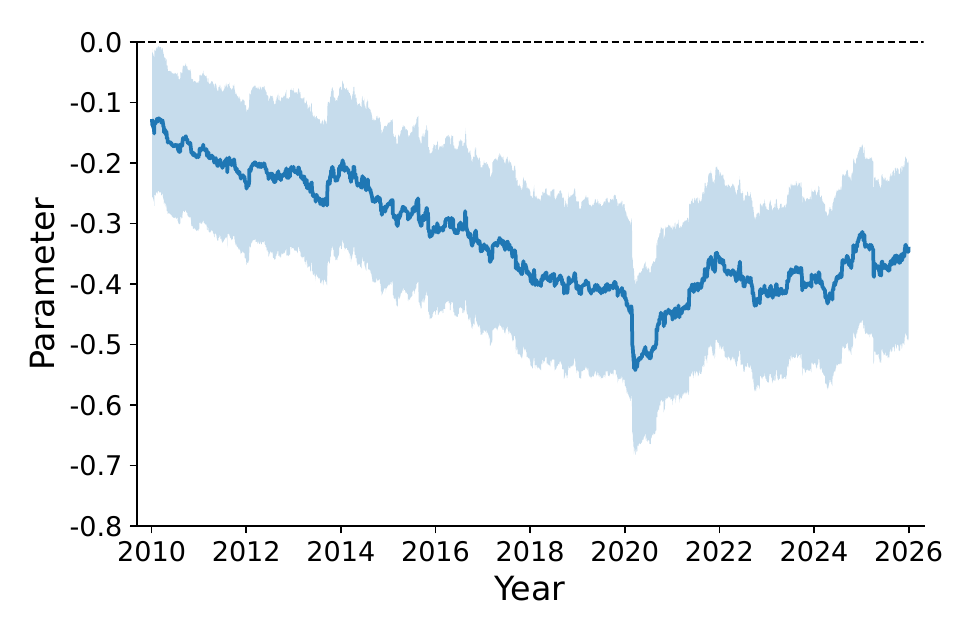}
    \includegraphics[width=0.495\textwidth]{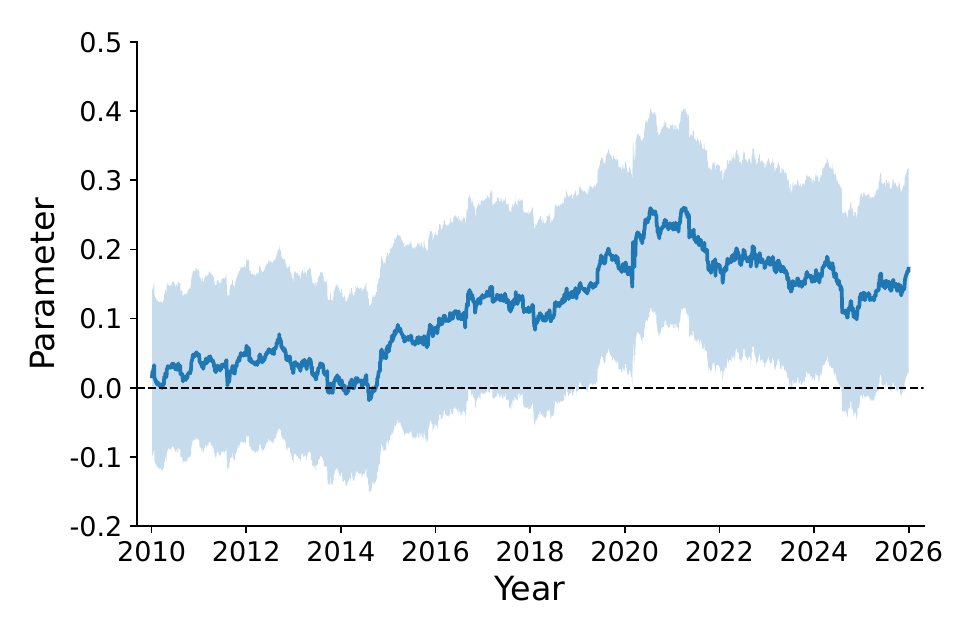}
    \includegraphics[width=0.495\textwidth]{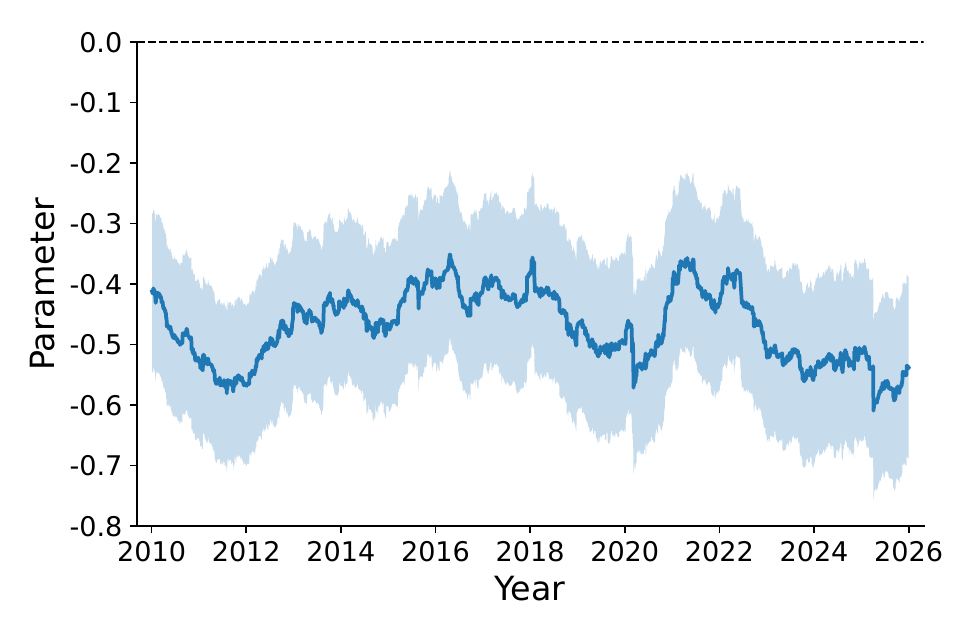}
    \includegraphics[width=0.495\textwidth]{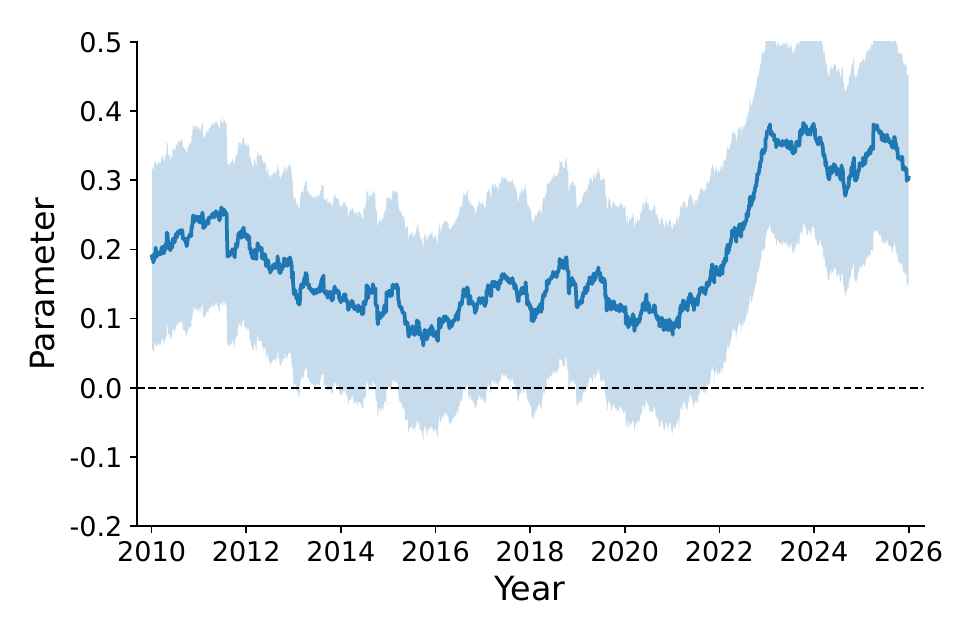}
  \caption{Filtered trajectories of 2 state vector elements $\btheta_t$ in the analysis of RVL-DLM for 3 selected sector ETFs.
  For each day, the figure shows the current posterior median and equal-tails 90\% credible interval for each coefficient.
  Rows show sectors XLB, XLU and XLE from top down. 
  {\em Left column:} contemporaneous RV coefficient on $x_t$, generally negative;
  {\em right column:} lag$-1$ RV coefficient on $x_{t-1}$, generally positive.
  The zero line is indicated.}
  \label{fig:RVL3trajectories}
\end{figure}

\begin{figure}[p!]
  \centering
    \includegraphics[width=0.495\textwidth]{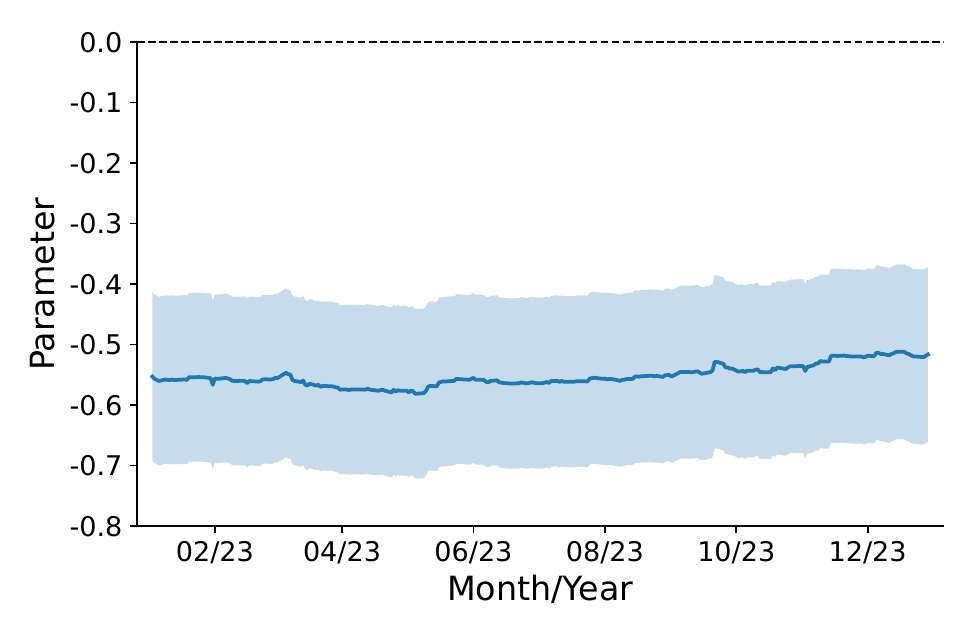}
    \includegraphics[width=0.495\textwidth]{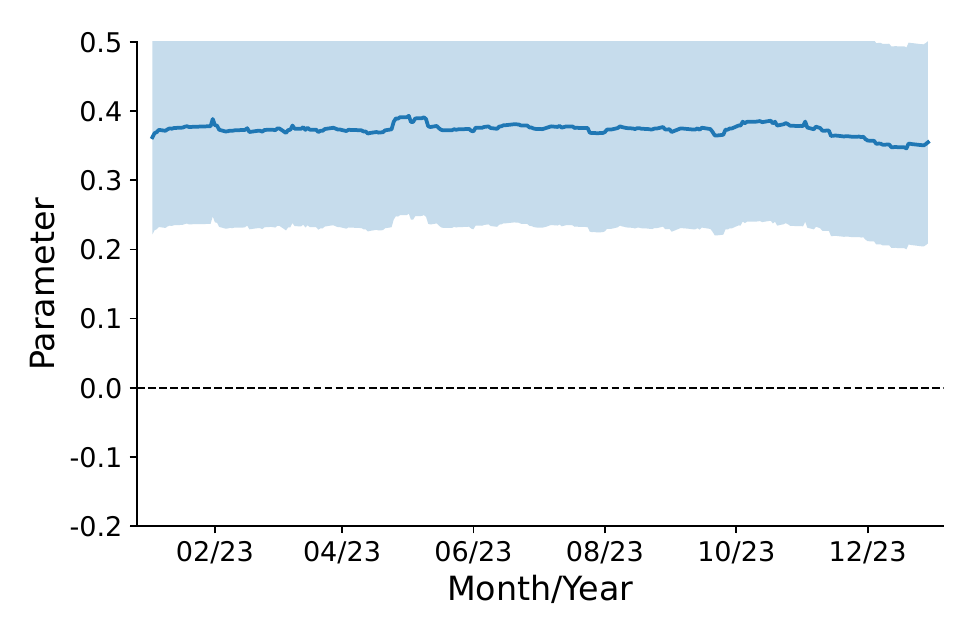}
    \includegraphics[width=0.495\textwidth]{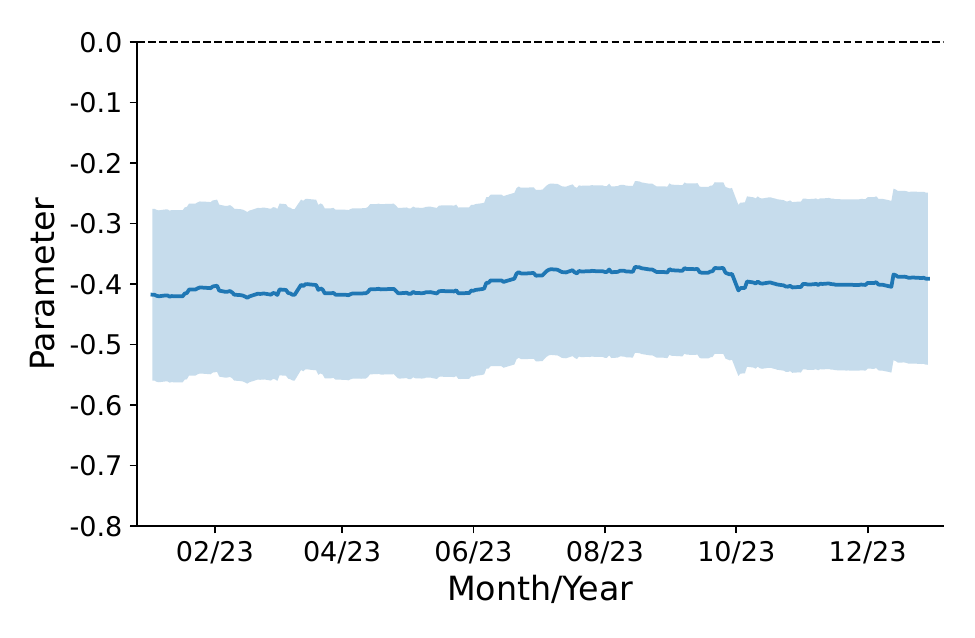}
    \includegraphics[width=0.495\textwidth]{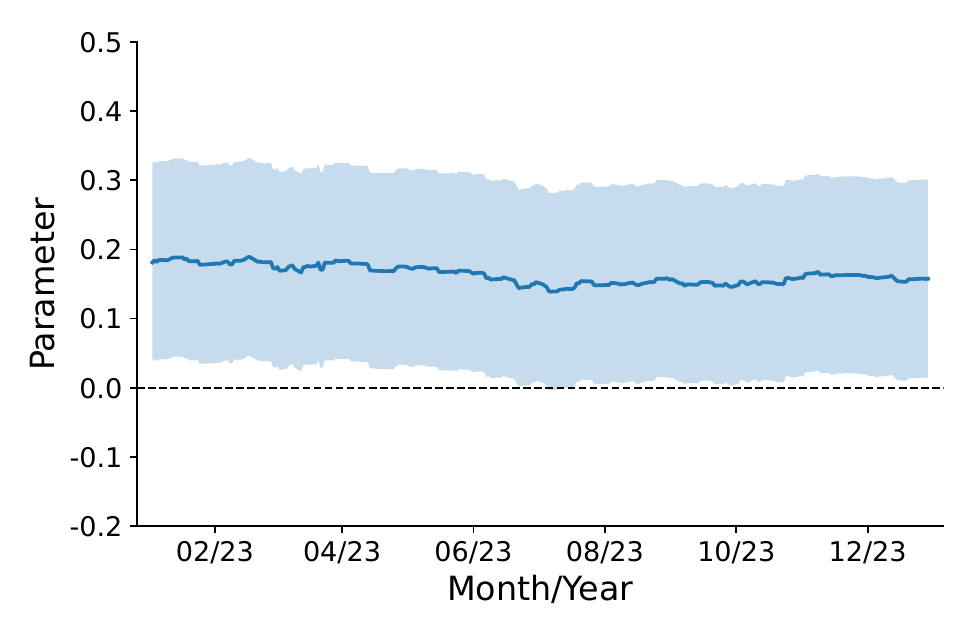}
    \includegraphics[width=0.495\textwidth]{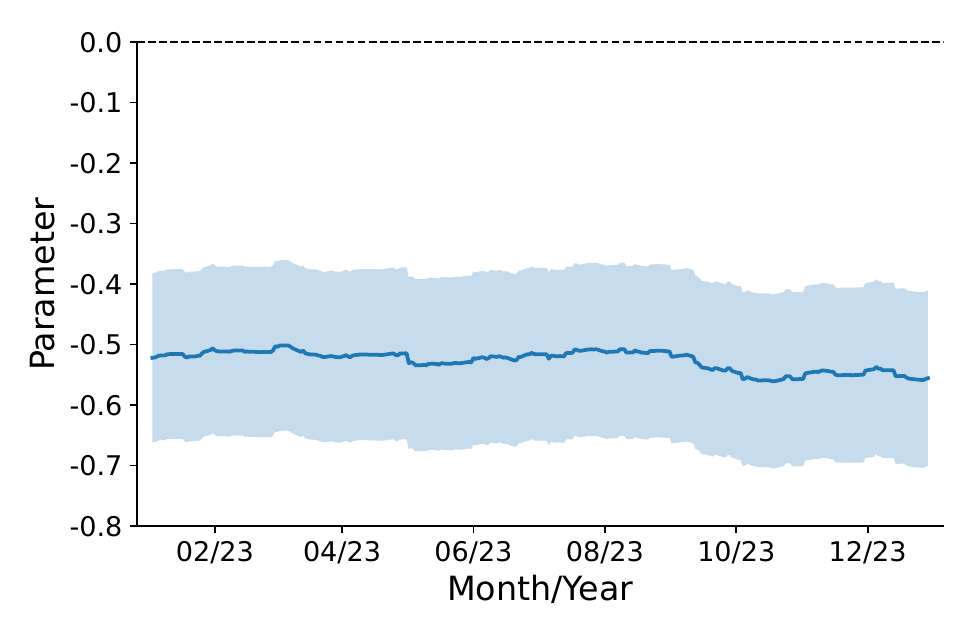}
    \includegraphics[width=0.495\textwidth]{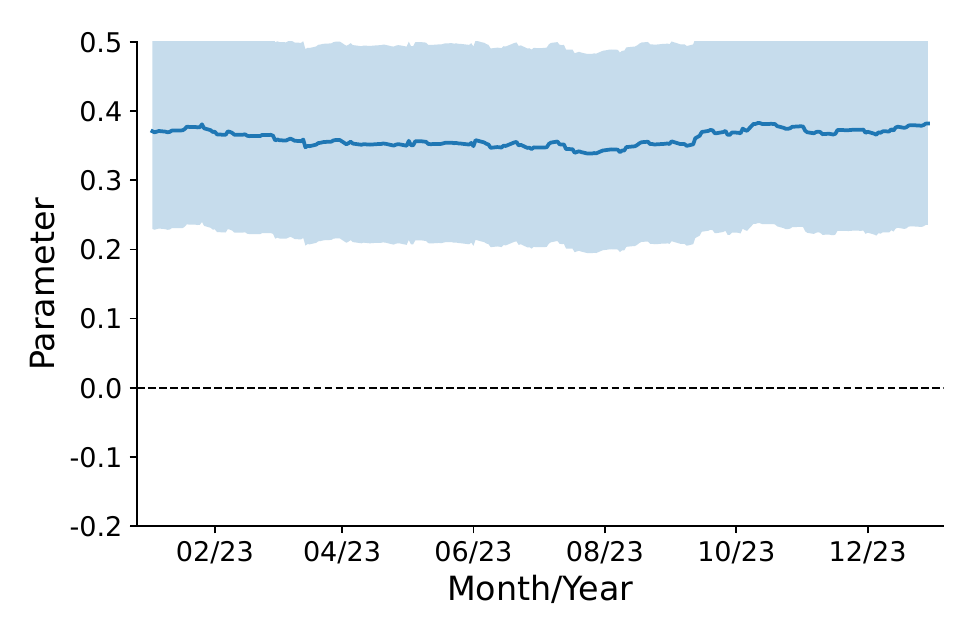}
  \caption{As in Figure~\ref{fig:RVL3trajectories} but now zoomed in to the one year period of 2023 data only.
  Rows show sectors XLB, XLU and XLE from top down. 
  {\em Left column:} contemporaneous RV coefficient on $x_t$;
  {\em right column:}  lag$-1$ RV coefficient on $x_{t-1}$.}
  \label{fig:RVL3trajectoriesZoomed}
\end{figure}

Some summaries of sequential inferences under the RVL-DLM analysis of each of 3 ETFs are shown in
Figures~\ref{fig:RVL3trajectories} and~\ref{fig:RVL3trajectoriesZoomed}. The former shows trajectories of posteriors for two elements of $\btheta_t$ 
 over the full evaluation period while the latter zooms-in to just the trading days in 2023.     The state vector elements are the time-varying coefficients on 
 $x_{t-1}$ and $x_t$, which summarize lagged and contemporaneous RV effects in the price equation rather than define a structural decomposition of volatility leverage and feedback.  In Figures~\ref{fig:RVL3trajectories} and~\ref{fig:RVL3trajectoriesZoomed}, the left-column panels show the contemporaneous coefficient on $x_t$, which is generally negative, while the right-column panels show the lag$-1$ coefficient on $x_{t-1}$, which is generally positive.  Equivalently, in these examples, the posterior medians for the lag$-1$ RV coefficient are generally positive, while those for the contemporaneous RV coefficient are generally negative.  Within the RVL-DLM specification, elevated lagged realized SD contributes a positive short-run term, whereas elevated current realized SD is associated with a negative contemporaneous adjustment to the log-price nowcast.  The latter negative contemporaneous association is in line with the high-frequency return--volatility asymmetries emphasized by \citet{BollerslevLitvinovaTauchen2006}, while the positive lagged term reflects the ETF-specific reduced-form daily price equation here. The figures also indicate the generally negative posterior relationship between the two coefficients (evident and quantified in the full posterior at each time, not shown here), with implied positive dependence in the absolute values of these coefficients over time. 
We note that, in the RV-DLM model that does not have the contemporaneous predictor $x_t,$ trajectories of inferences on the coefficient of the lagged predictor $x_{t-1}$ tend to vary around zero and at much lower values than indicated in the more general model, though with meaningful excursions from time to time (results not shown here).  This is an instance of the role of $x_{t-1}$ being unclear and often \lq\lq apparently'' insignificant in the simpler RV-DLM model, while adding the conditioning on $x_t$ via the RVL-DLM emphasizes its practical role consistent with econometric intuition.

  The additional Figure~\ref{fig:RVL3trajectoriesZoomed} simply focuses on a shorter 250 trading day period. The global  Figure~\ref{fig:RVL3trajectories} shows material fluctuations over  the weeks and months in these state vector elements, while looking at much shorter periods of time makes clear the rather stable evolutions inferred. This is important in understanding predictive value, as coefficients that vary wildly typically 
indicate poor potential for predictive value. On a short-term basis, these coefficients are clearly very stable indeed, contributing to the improvements seen in short-term forecasting for some of the ETFs.

\begin{figure}[ht!]
  \centering
    \includegraphics[width=0.495\textwidth]{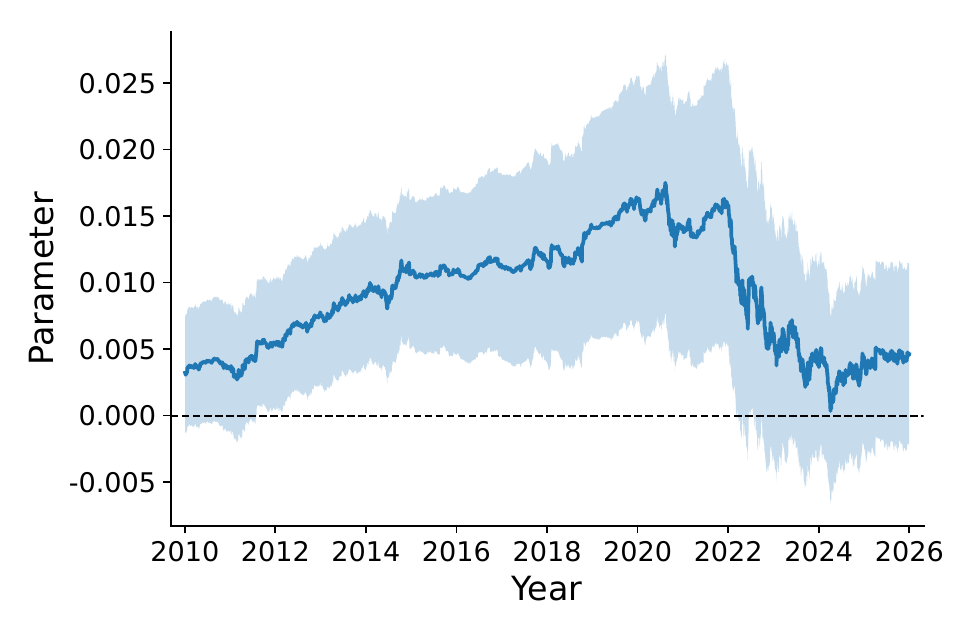}
    \includegraphics[width=0.495\textwidth]{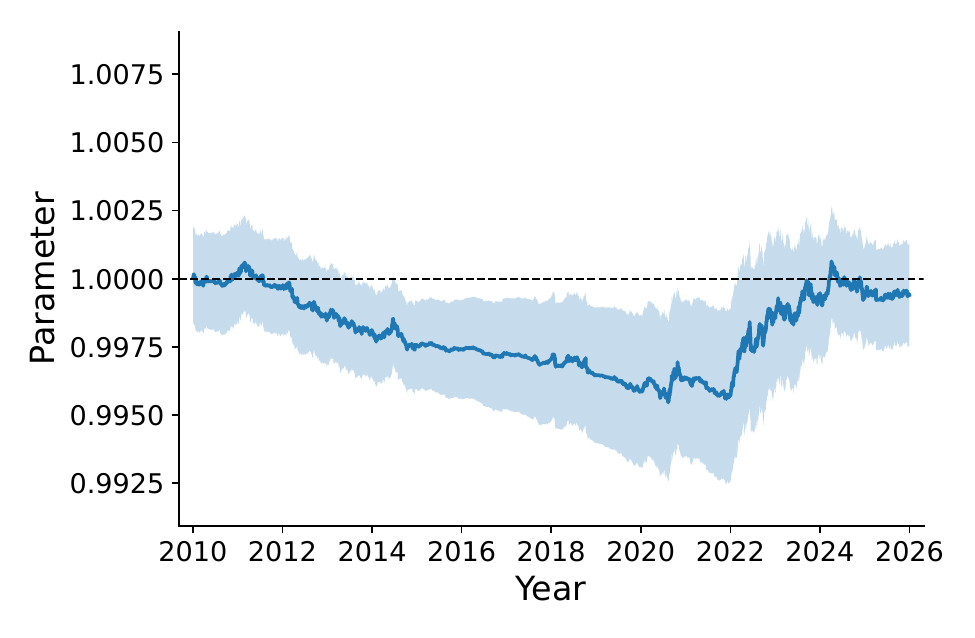}
    \includegraphics[width=0.495\textwidth]{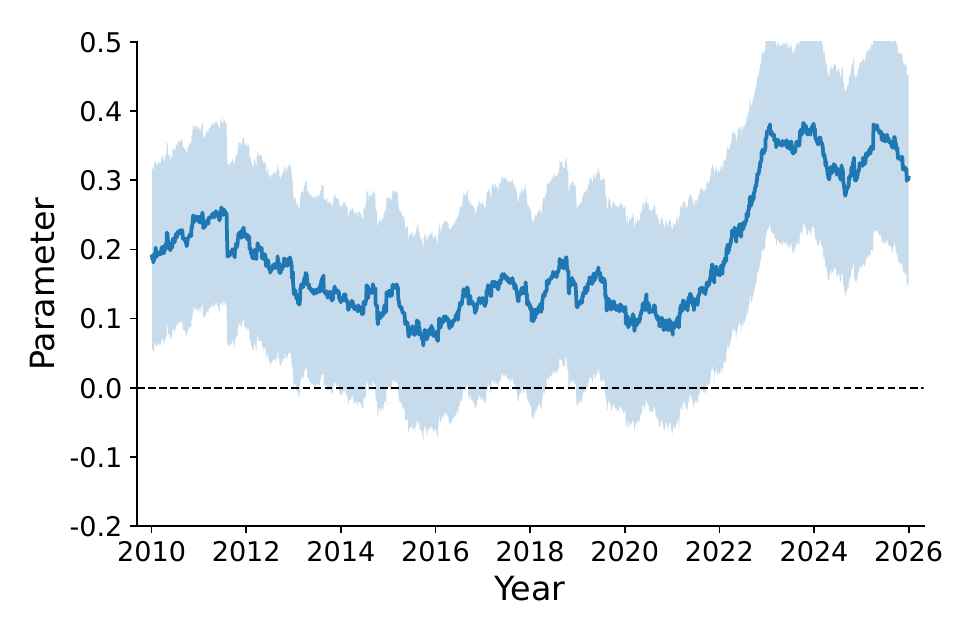}
    \includegraphics[width=0.495\textwidth]{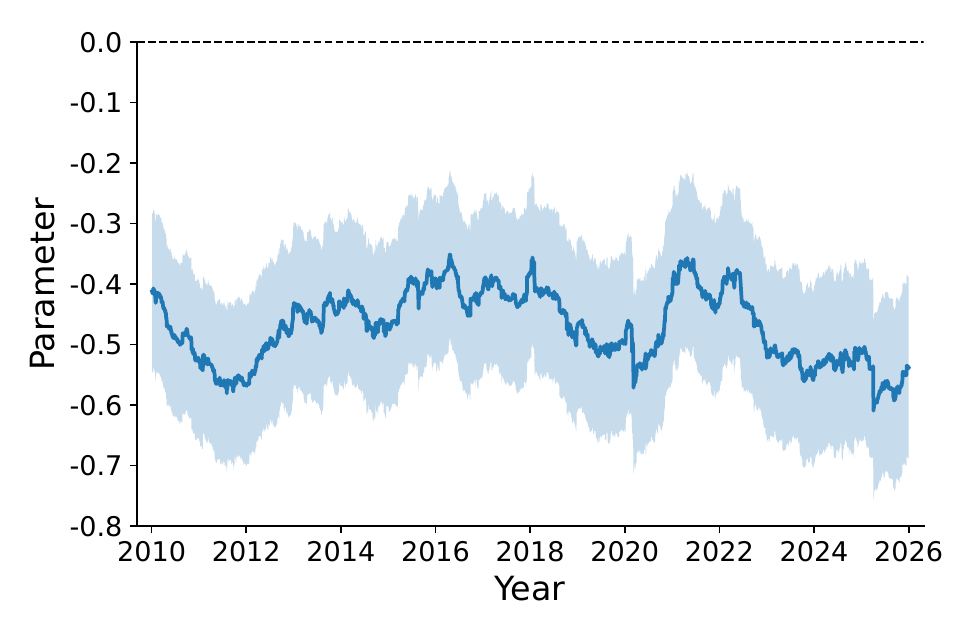}
  \caption{Filtered trajectories of all 4 elements of the state vector $\btheta_t$ in the analysis of RVL-DLM for the XLE series.
  At each day this shows the current posterior median and equal-tails 90\% credible interval for each coefficient.
  {\em Upper left:} local intercept; {\em upper right:} AR(1) log price coefficient;
  {\em lower left:} lag$-1$ RV coefficient on $x_{t-1}$, generally positive; {\em lower right:} contemporaneous RV coefficient on $x_t$, generally negative.
  The zero line is indicated.}
  \label{fig:XLEcoeftrajectories}
\end{figure}

 Figure~\ref{fig:XLEcoeftrajectories} shows the corresponding trajectories for the ETF series XLE over the full evaluation period. This now includes trajectories of the other 2 elements of the state-vector:  the TVAR(1) local intercept and its coefficient on lagged log price. Again this highlights the positive lagged RV effect of $x_{t-1}$ and the negative contemporaneous RV effect of $x_t$ in the model that includes both measures. The TVAR(1) coefficient is, as expected, inferred as
generally stable over time and very close to 1.  A value less than 1 is consistent with \lq\lq local stationarity" as a first-order feature of the model, and this is supported.  As something of an aside but worth noting, a value of precisely 1 for this coefficient implies that the model reduces to one for the 
return $y_t-y_{t-1}$ predicted by the local level and realized volatility elements; the inferred trajectory indicates evidence against this as a norm, supporting our general view that modeling (log) prices directly provides predictive value relative to models of returns directly.  Then, the local intercept term accounts partially for other aspects of local drifts in log prices. The 
role of the latter in adjusting model forecasts at the start of the Covid-19 pandemic in early 2020 is perhaps particularly notable. 
 
\newpage

\begin{figure}[t!]
  \centering
    \includegraphics[width=0.495\textwidth]{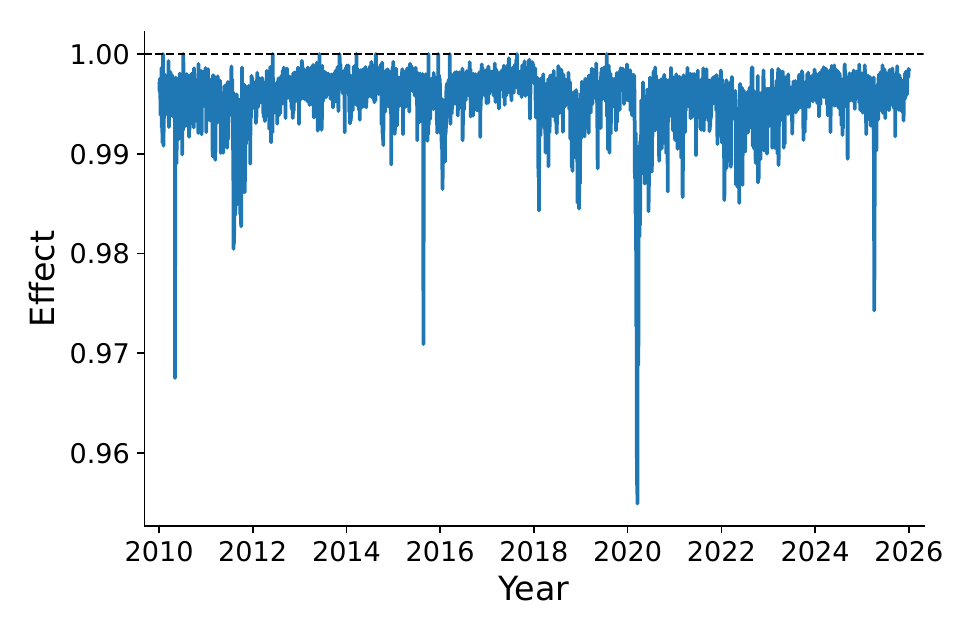}
    \includegraphics[width=0.495\textwidth]{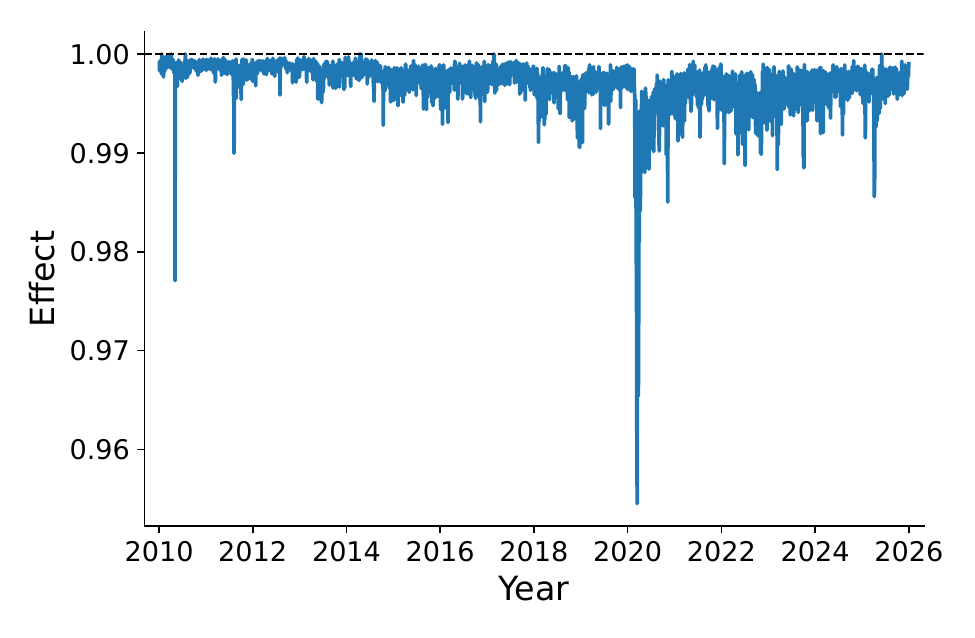}
    \includegraphics[width=0.495\textwidth]{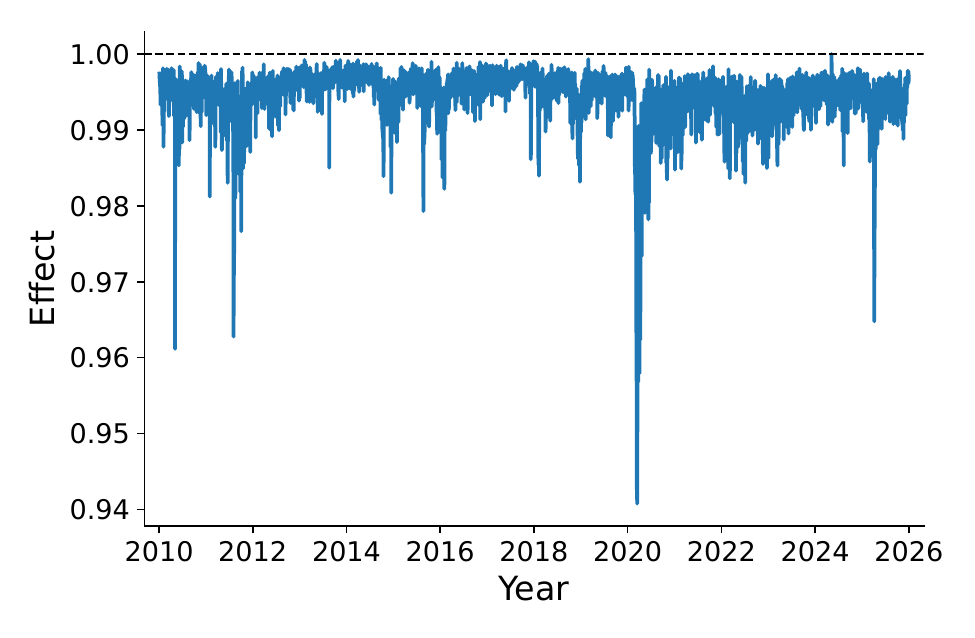}
  \caption{Trajectories of filtered posterior medians of the RVL-DLM contemporaneous regression effect of  $x_t$, converted to the  price scale for each of 3 sector ETFs.
  ETF sectors are: {\em upper left:} XLB; {\em upper right:} XLU; {\em lower:} XLE.
  The dashed line at 1 is consistent with no effect; values below 1 reflect the generally negative contemporaneous RV coefficient.}
  \label{fig:effectsonprice2ETFs}
\end{figure}

Addressing the practical impact of the contemporaneous RV measure $x_t,$
Figure~\ref{fig:effectsonprice2ETFs} shows trajectories of posterior medians of the
regression effect transformed to the price scale for ETFs XLB, XLU and XLE. Patterns for other ETFs are very similar. At the posterior median on each day $t$ of the state-vector element $\theta_{0x,t}$ for contemporaneous RV predictor $x_t,$ the estimated multiplicative effect is $\exp(\theta_{0x,t}x_t).$  Since the contemporaneous coefficient is negative in these examples, this component reduces the price-scale nowcast relative to the same model with the contemporaneous term set to zero.  The effect is generally close to 1, typically of the order of $1{-}2\%$ on the price scale, while there are evident downward spikes in periods of heightened market stress including the onset of the Covid-19 pandemic at the start of 2020 and the roll-out of so-called \lq\lq Trump tariffs'' in spring 2025. These contemporaneous adjustments should not be interpreted in isolation: the net realized volatility contribution also includes the positive lagged RV term, the local intercept, the AR(1) term and the volatility-dependent predictive scale. Other such events appear at different times that differentially impact subsets of ETFs, being more notable in some sectors than others. One such period in which S\&P sectors suffered losses and increased volatility was the spring 2010  \lq\lq flash crash", driven primarily by Euro debt crisis uncertainty and automated selling. This is evident in the examples shown in the figure. 
A more severe event in August 2015 was driven mainly by concerns about global economic slowdown and a resulting major drop in oil price; this led to significant selling of ETFs/ETPs, strongly evident in several sector ETFs including XLB as shown, 
though not so substantially in others.

\section{Summary Comments}
\label{sec:conclusion}

The new class of RVL-DLMs jointly models asset prices and realized volatility measures in a fully conjugate  Bayesian state-space framework. 
This integration of a new dynamic gamma model for realized variances with traditional SV-DLMs  for asset prices (log prices, or returns) can improve 
 volatility tracking and short-term prediction.   The theory yields analytic tractability for filtering and forecasting
 of realized volatilities and prices jointly.  Developments of the DLM component 
 are motivated by high-frequency econometrics and reduced-form contemporaneous and lagged price-volatility dependence.
The conditional mean of (log) prices 
may depend on both contemporaneous and lagged realized volatility proxies. The model then 
incorporates reduced-form volatility leverage and feedback  effects directly within the state-space structure, while maintaining analytic tractability.
 
The case study comparisons show strong evidence of improved one-step predictive performance of two OHLC-based model variants relative to 
the traditional univariate SV-DLM for asset prices.  The RV-DLM includes the lag$-1$ value of OHLC-based realized SD as a predictor in the 
price DLM, while the RVL-DLM also includes the contemporaneous value.   Including the latter without the former has been explored but to little effect. 
Comparisons with SV-DLM are fair in first extending the traditional model to include the same lagged realized volatility proxy as used in the RV-DLMs.  
The out-of-sample predictive improvements of the two bivariate models are systematic across all 10 series-- the 9 sector ETFs and the S\&P Index itself. 
Marginal likelihood predictive scores show substantial improvement gains that accumulate steadily over time through the evaluation period.
 
In addition to their real predictive value, the new models improve tracking and monitoring of latent volatility, highlight and directly quantify the roles and impact of realized volatility predictors in the price DLM, and offer an opportunity to understand reduced-form volatility leverage and feedback effects.  The empirical results further clarify the role of the contemporaneous realized-volatility term.  The magnitude of this incremental gain is sector-specific, but the sign of the cumulative log predictive advantage is uniformly positive. This supports using the RVL-DLM as the routine specification when contemporaneous realized volatility is available, with the RV-DLM retained as a simpler baseline and diagnostic comparison. The gain appears to arise less from large changes in point forecasts than from sharper predictive distributions and more efficient use of the realized-volatility information. The overall results are consistent with the higher information content of high-frequency realized volatility measures. In all cases, the RVL-DLM sharpens predictive densities for daily prices at negligible additional computational cost, making them attractive for routine forecasting and risk-control systems. 
 
Beyond the scope of the example in the current paper,  we note that the closed-form analysis for filtering-- coupled with easy direct simulation for one-step 
forecasting-- will underlie similarly trivial computations for multi-step prediction of the coupled price and realized volatility series. 
In complement to this, the new model also inherits the retrospective-analysis tools that underlie smoothing and backward sampling in the traditional SV-DLM, with the RV information entering through the filtered gamma parameters described above~\citep[][chap.~4]{PradoFerreiraWest2021}.  Further, the present model assumes a constant shape index $\alpha$ in the RV component, but this can be generalized to a predictable time-varying index $\alpha_t$ when the measurement precision of $z_t$ is expected to change over time.  One natural approach is to link $\alpha_t$ to rolling marginal likelihood diagnostics or to high-frequency estimates of realized quarticity and related second-order volatility variation, treating $\alpha_t$ as an adaptive precision parameter for the RV proxy.

For next steps in evaluating and extending the new models to potentially improve risk characterization in portfolio construction and management, extensions of the new models to multivariate settings are opened up in several directions.  First, the conditionally conjugate gamma:scaled-F construction we have presented  extends naturally to multivariate settings involving Wishart and multivariate-F distributions. With appropriate construction of empirical realized variance-covariance matrices based on high-frequency data, this opens potential to extend traditional Bayesian vector and matrix 
DLMs~(e.g.,~\citealp{WestHarrison1997}, chap.~16; \citealp{PradoFerreiraWest2021}, chap.~10)
while maintaining analytic tractability and computational simplicity. 
Second, there is immediate interest in embedding sets of bivariate RV-DLMs in more flexible and parsimonious multivariate models, especially simultaneous graphical dynamic linear models~\citep[e.g.][]{GruberWest2016,GruberWest2017,IrieWest2019portfoliosBA} that exploit the decouple/recouple approach to analysis based on inherent conditional independence structures~\citep{West2020Akaike}.  With the S\&P sector ETFs as an example setting, such a multivariate model would just couple the individual ETF RV-DLMs through assumed cross-series structures that still allow for fast and efficient sequential analysis and forecasting computations.  Such extensions 
would support real-time portfolio allocation, risk parity, and volatility-targeting strategies that are driven by easily simulated predictive distributions. 
These directions position RV-DLMs as building blocks for a broader class of Bayesian risk evaluation and trading models that make systematic use of realized volatility measures in a state-space framework. 

\subsubsection*{Data and code availability} 
Custom python code for data download (from \href{https://finance.yahoo.com/}{Yahoo Finance}) and preprocessing, together with all model analyses and figures in the paper, is publicly available at 
\if0\blind 
     the github repository
    \href{https://github.com/PSW1998/rv-dlm-paper-figures}{github.com/PSW1998/rv-dlm-paper-figure}.
    \small \setstretch{1.0}      
\else
    a githib repository.
\fi

\if0\blind {\small \setstretch{1.0}} \fi

\bibliography{RVDLM}

@Article{WatanableNakajima2024,
  author  = {Toshiaki Watanabe and Jouchi Nakajima},
  journal = {Journal of Empirical Finance},
  title   = {High-frequency realized stochastic volatility model},
  year    = {2024},
  issn    = {0927--5398},
  pages   = {101559},
  volume  = {79},
  doi     = {https://doi.org/10.1016/j.jempfin.2024.101559},
}

@article{BollerslevLitvinovaTauchen2006,
  author  = {Bollerslev, Tim and Litvinova, Julia and Tauchen, George},
  title   = {Leverage and Volatility Feedback Effects in High-Frequency Data},
  journal = {Journal of Financial Econometrics},
  year    = {2006},
  volume  = {4},
  number  = {3},
  pages   = {353--384},
  doi     = {10.1093/jjfinec/nbj014}
}

@Article{TakahashiWatanabeOmori2024,
  author  = {Makoto Takahashi and Toshiaki Watanabe and Yasuhiro Omori},
  journal = {Econometrics and Statistics},
  title   = {Forecasting Daily Volatility of Stock Price Index Using Daily Returns and Realized Volatility},
  year    = {2024},
  issn    = {2452-3062},
  pages   = {34--56},
  volume  = {32},
  doi     = {https://doi.org/10.1016/j.ecosta.2021.08.002},
}

@Article{IrieWest2019portfoliosBA,
  author  = {K. Irie and M. West},
  journal = {Bayesian Analysis},
  title   = {Bayesian emulation for multi-step optimization in decision problems},
  year    = {2019},
  note    = {Published online April 19, 2018},
  pages   = {137--160},
  volume  = {14},
  number = {1},
  arxiv   = {1607.01631},
  doi     = {10.1214/18-BA1105},
  url     = {https://projecteuclid.org/euclid.ba/1524103230},
}

@Article{West1993a,
  author       = {M. West},
  journal      = {Computing Science and Statistics},
  title        = {Mixture models, {M}onte {C}arlo, {B}ayesian updating and dynamic models},
  year         = {1993},
  pages        = {325--333},
  volume       = {24},
  creationdate = {2010-03-16T00:00:00},
  key          = {fds29244},
  url          = {http://www.stat.duke.edu/~mw/MWextrapubs/West1993a.pdf},
}

@Book{Taylor1986,
  author    = {Taylor, Stephen J.},
  publisher = {John Wiley \& Sons},
  title     = {Modelling Financial Time Series},
  year      = {1986},
  address   = {Chichester},
}

@Article{JacquierPolsonRossi1994,
  author  = {Jacquier, Eric and Polson, Nicholas G. and Rossi, Peter E.},
  journal = {Journal of Business \& Economic Statistics},
  title   = {Bayesian Analysis of Stochastic Volatility Models},
  year    = {1994},
  number  = {4},
  pages   = {371--389},
  volume  = {12},
  doi     = {10.1080/07350015.1994.10524553},
}

@InCollection{Shephard1996,
  author    = {Shephard, Neil},
  booktitle = {Time Series Models in Econometrics, Finance and Other Fields},
  publisher = {Chapman \& Hall},
  title     = {Statistical Aspects of ARCH and Stochastic Volatility},
  year      = {1996},
  address   = {London},
  editor    = {Cox, David R. and Hinkley, David V. and Barndorff-Nielsen, Ole E.},
  pages     = {1--67},
}

@Article{BarndorffNielsenHansenLundeShephard2008,
  author  = {Barndorff-Nielsen, Ole E. and Hansen, Peter R. and Lunde, Asger and Shephard, Neil},
  journal = {Econometrica},
  title   = {Designing Realized Kernels to Measure the Ex-Post Variation of Equity Prices in the Presence of Noise},
  year    = {2008},
  number  = {6},
  pages   = {1481--1536},
  volume  = {76},
  doi     = {10.3982/ECTA6495},
}

@Article{KoopmanScharth2013,
  author  = {Koopman, Siem Jan and Scharth, Marcel},
  journal = {Journal of Financial Econometrics},
  title   = {The Analysis of Stochastic Volatility in the Presence of Daily Realized Measures},
  year    = {2013},
  number  = {1},
  pages   = {76--115},
  volume  = {11},
  doi     = {10.1093/jjfinec/nbs016},
}

@Article{Engle2002,
  author  = {Engle, Robert F.},
  journal = {Journal of Applied Econometrics},
  title   = {New Frontiers for {ARCH} Models},
  year    = {2002},
  number  = {5},
  pages   = {425--446},
  volume  = {17},
  doi     = {10.1002/jae.683},
}

@Article{EngleGallo2006,
  author  = {Engle, Robert F. and Gallo, Giampiero M.},
  journal = {Journal of Econometrics},
  title   = {A Multiple Indicators Model for Volatility Using Intra-Daily Data},
  year    = {2006},
  number  = {1--2},
  pages   = {3--27},
  volume  = {131},
  doi     = {10.1016/j.jeconom.2005.01.018},
}

@Article{ShephardSheppard2010,
  author  = {Shephard, Neil and Sheppard, Kevin},
  journal = {Journal of Applied Econometrics},
  title   = {Realising the Future: Forecasting with High-Frequency-Based Volatility ({HEAVY}) Models},
  year    = {2010},
  number  = {2},
  pages   = {197--231},
  volume  = {25},
  doi     = {10.1002/jae.1158},
}

@Article{PittShephard1999,
  author  = {Pitt, Michael K. and Shephard, Neil},
  journal = {Journal of the American Statistical Association},
  title   = {Filtering via Simulation: Auxiliary Particle Filters},
  year    = {1999},
  number  = {446},
  pages   = {590--599},
  volume  = {94},
  doi     = {10.1080/01621459.1999.10474153},
}

@Book{PradoFerreiraWest2021,
  author    = {Prado, Raquel and Ferreira, Marco A. R. and West, Mike},
  publisher = {Chapman and Hall/CRC},
  title     = {Time Series: Modeling, Computation, and Inference},
  year      = {2021},
  address   = {Boca Raton},
  edition   = {2},
  doi       = {10.1201/9781351259422},
}

@inproceedings{black_1976,
  author  = {Black, Fischer},
  booktitle = {Proceedings of the American Statistical Association, Business and Economic Statistics Section},
  title   = {Studies of Stock Price Volatility Changes},
  year    = {1976},
  pages   = {177--181},
}

@Article{NAKAJIMA20123690,
  author   = {Jouchi Nakajima and Yasuhiro Omori},
  journal  = {Computational Statistics \& Data Analysis},
  title    = {Stochastic volatility model with leverage and asymmetrically heavy-tailed error using {GH} skew {S}tudent’s t-distribution},
  year     = {2012},
  issn     = {0167-9473},
  number   = {11},
  pages    = {3690-3704},
  volume   = {56},
  doi      = {https://doi.org/10.1016/j.csda.2010.07.012},
  url      = {https://www.sciencedirect.com/science/article/pii/S0167947310002859},
}

@Article{NAKAJIMA20092335,
  author   = {Jouchi Nakajima and Yasuhiro Omori},
  journal  = {Computational Statistics \& Data Analysis},
  title    = {Leverage, heavy-tails and correlated jumps in stochastic volatility models},
  year     = {2009},
  issn     = {0167-9473},
  number   = {6},
  pages    = {2335-2353},
  volume   = {53},
  doi      = {https://doi.org/10.1016/j.csda.2008.03.015},
  url      = {https://www.sciencedirect.com/science/article/pii/S0167947308001680},
}

@Article{OMORI2007425,
  author   = {Yasuhiro Omori and Siddhartha Chib and Neil Shephard and Jouchi Nakajima},
  journal  = {Journal of Econometrics},
  title    = {Stochastic volatility with leverage: Fast and efficient likelihood inference},
  year     = {2007},
  issn     = {0304-4076},
  number   = {2},
  pages    = {425-449},
  volume   = {140},
  doi      = {https://doi.org/10.1016/j.jeconom.2006.07.008},
  url      = {https://www.sciencedirect.com/science/article/pii/S0304407606001436},
}

@InCollection{Liu2001,
  author    = {Liu, Jane and West, Mike},
  editor    = {Doucet, Arnaud and de Freitas, Nando and Gordon, Neil},
  pages     = {197--223},
  publisher = {Springer New York},
  title     = {Combined Parameter and State Estimation in Simulation-Based Filtering},
  year      = {2001},
  address   = {New York, NY},
  isbn      = {978-1-4757-3437-9},
  booktitle = {Sequential Monte Carlo Methods in Practice},
  doi       = {10.1007/978-1-4757-3437-9_10},
  url       = {https://doi.org/10.1007/978-1-4757-3437-9_10},
}

@Article{engle_ng_1993,
  author  = {Engle, Robert F. and Ng, Victor K.},
  journal = {Journal of Finance},
  title   = {Measuring and Testing the Impact of News on Volatility},
  year    = {1993},
  number  = {5},
  pages   = {1749--1778},
  volume  = {48},
}

@Article{West2020Akaike,
  author  = {M. West},
  journal = {Annals of the Institute of Statistical Mathematics},
  title   = {Bayesian forecasting of multivariate time series: {S}calability, structure uncertainty and decisions (with discussion)},
  year    = {2020},
  pages   = {1--44},
  volume  = {72},
  number = {1}, 
  arxiv   = {1911.09656},
  doi     = {10.1007/s10463-019-00741-3},
}

@Article{GruberWest2016,
  author  = {L. F. Gruber and M. West},
  journal = {Bayesian Analysis},
  title   = {{GPU}-accelerated {B}ayesian learning and forecasting in simultaneous graphical dynamic linear models},
  year    = {2016},
  pages   = {125--149},
  volume  = {11},
  number = {1}, 
  doi     = {10.1214/15-BA946},
  url     = {http://projecteuclid.org/euclid.ba/1425304898},
}

@Article{GruberWest2017,
  author       = {L. F. Gruber and M. West},
  journal      = {Econometrics and Statistics},
  title        = {Bayesian forecasting and scalable multivariate volatility analysis using simultaneous graphical dynamic linear models},
  year         = {2017},
  pages        = {3--22},
  volume       = {3},
  arxiv        = {1606.08291},
  comment      = {Published online March 12, 2017},
  creationdate = {2017-09-22T00:00:00},
  doi          = {10.1016/j.ecosta.2017.03.003},
  url          = {http://www.sciencedirect.com/science/article/pii/S2452306217300163},
}

@Article{Aguilar2000,
  author       = {O. Aguilar and M. West},
  journal      = {Journal of Business and Economic Statistics},
  title        = {Bayesian dynamic factor models and portfolio allocation},
  year         = {2000},
  pages        = {338--357},
  volume       = {18},
  creationdate = {2010-03-16T00:00:00},
  key          = {fds29168},
    number  = {3},
  url          = {http://www.stat.duke.edu/~mw/MWextrapubs/Aguilar2000.pdf},
}

@InCollection{TallmanWest2024,
  author    = {E. Tallman and M. West},
  booktitle = {Recent Developments in Bayesian Econometrics and Their Applications},
  publisher = {Springer},
  title     = {Predictive decision synthesis for portfolios: {B}etting on better models},
  year      = {2024},
  editor    = {S. Mazur and P. {\"{O}}sterhol},
  note      = {arXiv:2405.01598},
  arxiv     = {2405.01598},
}

@Article{RogersSatchell1991,
  author  = {Rogers, Leonard C. G. and Satchell, Stephen E.},
  journal = {The Annals of Applied Probability},
  title   = {Estimating Variance from High, Low, and Closing Prices},
  year    = {1991},
  number  = {4},
  pages   = {504--512},
  volume  = {1},
}

@Article{KimShephardChib1998,
  author   = {Kim, Sangjoon and Shephard, Neil and Chib, Siddhartha},
  journal  = {The Review of Economic Studies},
  title    = {Stochastic volatility: {L}ikelihood inference and comparison with {ARCH} models},
  year     = {1998},
  issn     = {0034-6527},
  month    = {07},
  number   = {3},
  pages    = {361-393},
  volume   = {65},
  doi      = {10.1111/1467-937X.00050},
  eprint   = {https://academic.oup.com/restud/article-pdf/65/3/361/4459409/65-3-361.pdf},
  url      = {https://doi.org/10.1111/1467-937X.00050},
}

@Article{WestHarrison1986,
  author  = {M. West and P. J. Harrison},
  journal = {Journal of the American Statistical Association},
  title   = {Monitoring and adaptation in {B}ayesian forecasting models},
  year    = {1986},
  pages   = {741--750},
  volume  = {81},
  number ={395},
}

@Book{WestHarrison1997,
  author    = {West, Mike and Harrison, P. Jeff},
  publisher = {Springer},
  title     = {Bayesian Forecasting and Dynamic Models},
  year      = {1997},
  address   = {New York},
  edition   = {2},
  isbn      = {9780387947259},
}

@Article{HansenHuangShek2012,
  author  = {Hansen, Peter R. and Huang, Zhuo and Shek, Howard H.},
  journal = {Journal of Applied Econometrics},
  title   = {Realized {GARCH}: {A} joint model for returns and realized measures of volatility},
  year    = {2012},
  number  = {6},
  pages   = {877--906},
  volume  = {27},
  doi     = {10.1002/jae.1234},
}

@Article{TakahashiOmoriWatanabe2009,
  author  = {Takahashi, Makoto and Omori, Yasuhiro and Watanabe, Toshiaki},
  journal = {Computational Statistics \& Data Analysis},
  title   = {Estimating stochastic volatility models using daily returns and realized volatility simultaneously},
  year    = {2009},
  number  = {6},
  pages   = {2404--2426},
  volume  = {53},
  doi     = {10.1016/j.csda.2008.07.039},
}

@Article{ABDL2003,
  author  = {Andersen, Torben G. and Bollerslev, Tim and Diebold, Francis X. and Labys, Paul},
  journal = {Econometrica},
  title   = {Modeling and Forecasting Realized Volatility},
  year    = {2003}, 
  number  = {2},
  pages   = {579--625},
  volume  = {71},
  doi     = {10.1111/1468-0262.00418},
}

@Article{BNS2002,
  author  = {Barndorff-Nielsen, Ole E. and Shephard, Neil},
  journal = {Journal of the Royal Statistical Society: Series B},
  title   = {Econometric Analysis of Realized Volatility and Its Use in Estimating Stochastic Volatility Models},
  year    = {2002},
  number  = {2},
  pages   = {253--280},
  volume  = {64},
  doi     = {10.1111/1467-9868.00336},
}
\bibliographystyle{chicago} 

\end{document}